\documentclass[pra,preprintnumbers,amsmath,amssymb,showpacs,
nofootinbib,floatfix]{revtex4}

\usepackage{graphicx,bm,amsmath}

\makeatletter
\def\graphicscale{\twocolumn@sw{0.33}{0.4}}
\def\graphicthreescale{\twocolumn@sw{0.33}{0.4}}
\makeatother

\begin{document}

\title{Anisotropic perturbations in three-dimensional O($N$)-symmetric vector models}

\author{Martin Hasenbusch}
\affiliation{Institut f\"ur Physik, Humboldt-Universit\"at
zu Berlin, Newtonstr. 15, 12489 Berlin, Germany}
\email{Martin.Hasenbusch@physik.hu-berlin.de}
\author{Ettore Vicari}
\affiliation{Dipartimento di Fisica dell'Universit\`a di Pisa
        and INFN, I-56127 Pisa, Italy}
\email{Ettore.Vicari@df.unipi.it}

\begin{abstract}
  We investigate the effects of anisotropic perturbations in three-dimensional
  O($N$)-symmetric vector models. In order to assess their relevance for the
  critical behavior, we determine the renormalization-group dimensions of the
  anisotropic perturbations associated with the first few spin values of the
  representations of the O($N$) group, because the lowest spin values give
  rise to the most important effects.  In particular, we determine them up to
  spin 4 for $N=2,3,4$, by finite-size analyses of Monte Carlo simulations of
  lattice O($N$) models, achieving a significant improvement of their
  accuracy.  These results are relevant for several physical systems, such as
  density-wave systems, magnets with cubic symmetry, and multicritical
  phenomena arising from the competition of different order parameters.

\end{abstract}



\maketitle


\section{Introduction and Summary}
\label{intro}
Many continuous phase transitions observed in nature belong to the O($N$)
vector universality classes, which are characterized by an $N$-component order
parameter with O($N$) symmetry and the symmetry breaking
O($N$)$\rightarrow$O($N-1$).
The superfluid transition in $^4$He, the formation of Bose-Einstein
condensates, density wave systems, transitions in magnets with easy-plane
anisotropy, and in superconductors belong to the XY or O(2) universality
class; the Curie transition in isotropic magnets, zero-temperature quantum
transitions in two-dimensional antiferromagnets, are examples for the
Heisenberg or O(3) universality class; the O(4) universality class is relevant
for the finite-temperature transition in two-flavor quantum chromodynamics,
the theory of strong interactions.  See, e.g., Refs.~\cite{ZJ-book,PV-r} for
reviews.

In the absence of external fields, the phase transition of O($N$)-symmetric
vector models is driven by only one relevant parameter, which is usually
associated with the temperature.  The corresponding RG dimension is
$y_t=1/\nu$ where $\nu$ is the correlation-length exponent.  The leading odd
perturbation, which breaks the O($N$) symmetry, is associated with the
external field $h$ coupled to the order parameter; it has RG dimension
$y_h=(d+2-\eta)/2$, where $\eta$ is the exponent controlling the power-law
space-dependence of the two-point correlation function of the order parameter
at criticality. The asymptotic critical power-law behaviors of
O($N$)-symmetric vector models have been determined with high accuracy. In
Table~\ref{tabcexp} we report some of the most accurate estimates of the
critical exponents $\nu$ and $\eta$, and of the leading and next-to-leading
scaling-correction exponents $\omega$ and $\omega_2$, which characterize the
dominant corrections to the universal scaling.

\begin{table}[tbp]
\caption{Some of the most accurate results for the critical exponents
of the three-dimensional O($N$) vector universality classes with 
$N=2,3,4,5$.  We
report estimates of $\nu$ and $\eta$, and of the leading and
next-to-leading scaling correction exponents, obtained by lattice
techniques (LT) based on Monte Carlo simulations and/or
high-temperature expansions, and by quantum field theory (FT)
techniques such as high-order perturbative expansions.  The results
without reference have been obtained in this paper.
A more complete review of results can be found in Ref.~\cite{PV-r}.}
\label{tabcexp}
\begin{ruledtabular}
\begin{tabular}{clllll}
\multicolumn{1}{c}{$N$}& 
\multicolumn{1}{c}{method}& 
\multicolumn{1}{c}{$\nu$}&
\multicolumn{1}{c}{$\eta$}& 
\multicolumn{1}{c}{$\omega$}&
\multicolumn{1}{c}{$\omega_2$}\\ 
\colrule

2 & LT & 
0.6717(1) \cite{CHPV-06} & 0.0381(2) \cite{CHPV-06}
&  0.785(20) \cite{CHPV-06} & \\

  & FT & 
0.6703(15) \cite{GZ-98} & 0.0354(25) \cite{GZ-98}
&  0.789(11) \cite{GZ-98} & 1.77(7) \cite{NR-84} \\\hline

3 & LT & 
0.7112(5) \cite{CHPRV-02}& 0.0375(5) \cite{CHPRV-02}& 
0.773 \cite{H-01} & \\

  & & 0.7117(5) \cite{this} & 0.0378(5) \cite{this} & & \\

  & & 0.7116(10) & 0.0378(3) & & \\

  & FT & 0.7073(35) \cite{GZ-98} & 0.0355(25) \cite{GZ-98}
&  0.782(13) \cite{GZ-98} & 1.78(11) \cite{NR-84} \\ \hline

4 & LT & 
0.749(2) \cite{H-01} & 0.0365(10) \cite{H-01}& 0.765 \cite{H-01} & \\

& & 0.7477(8) \cite{Deng-06} & 0.0360(4) \cite{Deng-06} & & \\

  & &  0.750(2) & 0.0360(3) & & \\

  & FT & 0.741(6) \cite{GZ-98} & 0.0350(45) \cite{GZ-98}
&  0.774(20) \cite{GZ-98} & \\\hline

5 & LT & 
0.779(3) \cite{HPV-05} & 0.034(1) \cite{HPV-05}& & \\

  & FT & 0.762(7) \cite{CPV-03} & 0.034(4) \cite{CPV-03}
&  0.790(15) \cite{CPV-03} & \\

\end{tabular}
\end{ruledtabular}
\end{table}

In this paper we study the effects of anisotropic perturbations breaking the
O($N$) symmetry, which cannot be related to an external vector field coupled
to the order parameter, but which are represented by composite operators with
more complex transformation properties under the O($N$) group.  An interesting
question is whether they change the critical behavior, or whether they do not
affect it so that the symmetry shown by the critical correlations is larger
than that of the microscopic model.  This issue arises in several physical
contexts.  Anisotropy in magnetic systems may naturally arise due to the cubic
structure of the underlying lattice, giving rise to anisotropic interactions
terms, see, e.g., Ref.~\cite{Aharony-76}.  The relevance of the anisotropic
perturbations determines also the nature of the multicritical behavior at the
meeting point of two transition lines with different O($n_1$) and O($n_2$)
symmetries, in particular, whether the symmetry gets effectively enlarged to
$O(n_1+n_2)$, see, e.g., Refs.~\cite{FN-74,NKF-74,CPV-03}.  Another
interesting issue is the critical behavior of secondary order parameters,
which are generally represented by powers of the order parameter transforming
as higher representations of the O($N$) group; their critical behaviors can be
measured in density wave systems, such as liquid
crystals~\cite{Brock-etal-86,ABBL-86,Aharony-etal-95}, see also
Refs.~\cite{AM-83,HHTG-95,ZMHCLGGS-96,Bak-80}.

Let us consider the general problem of the O($N$)-symmetric theory in the
presence of an external field $h_p$ coupled to a perturbation $P$.
Assuming $P$ to be an eigenoperator of the RG transformations, the singular
part of the free energy for the reduced temperature $t\to 0$ and $h_p
\to 0$ can be written as
\begin{equation}
F_{\rm sing} =      |t|^{d\nu} 
f\left( h_p/|t|^{y_p\nu} \right),
\label{freeen}
\end{equation}
where $y_p$ is the RG dimension of $h_p$, and $f(x)$ is a scaling function.
Therefore, the RG dimensions of the anisotropic external fields
quantitatively control their capability to influence or change the asymptotic
critical behavior when $y_p>0$.

In the field-theoretical (FT) 
framework the O($N$)-symmetric vector model is represented by
the O($N$)-symmetric Landau-Ginzburg-Wilson theory
\begin{equation}
{\cal H} = \int d^d x \left[
\case{1}{2} (\partial_\mu\Phi)^2  + \case{1}{2} r \Phi^2
+ \case{1}{4!} u (\Phi^2)^2 + h\cdot\Phi \right],
\label{phi4h}
\end{equation}
where $\Phi$ is an $N$-component real field and $h$ an external field.  The
anisotropic perturbations are conveniently classified~\cite{Wegner-76,PV-r}    
using irreducible representations of the O($N$) internal group,
characterized by the spin value $l$.
Let us consider the perturbation $P_{m,l}$ defined by the power $m$ of
the order parameter and the spin representation $l$ of the O($N$) group
\begin{equation}
    P_{m,l}^{a_1...a_l}(\Phi) = (\Phi^2)^{(m-l)/2} Q_l^{a_1...a_l}(\Phi)
\label{pmldef}
\end{equation}
where $Q_l^{a_1...a_l}$ is a homogeneous polynomial of degree $l$ 
that is symmetric and traceless in the $l$ indices:
\begin{eqnarray}
&&Q^{a}_{1}(\Phi) = \Phi^a 
 \label{spin1}\\
&&Q^{ab}_{2}(\Phi) = \Phi^a \Phi^b - {1\over N} \delta^{ab} \Phi^2
 \label{spin2}\\
&&Q^{abc}_3(\Phi) = \Phi^a \Phi^b \Phi^c - {\Phi^2\over N + 2} \left(\Phi^a
  \delta^{bc} + \Phi^b \delta^{ac} + \Phi^c \delta^{ab}\right)
 \label{spin3}\\
&&Q^{abcd}_{4}(\Phi) = \Phi^a \Phi^b \Phi^c \Phi^d  
- {1\over N+4} \Phi^2 \left( 
        \delta^{ab} \Phi^c \Phi^d + \delta^{ac} \Phi^b \Phi^d + 
        \delta^{ad} \Phi^b \Phi^c + \delta^{bc} \Phi^a \Phi^d + 
        \delta^{bd} \Phi^a \Phi^c + \delta^{cd} \Phi^a \Phi^b \right) 
\nonumber \\
   && \qquad + {1\over (N+2)(N+4)} (\Phi^2)^2 \left(
         \delta^{ab} \delta^{cd} + \delta^{ac} \delta^{bd} + 
         \delta^{ad} \delta^{bc} \right)
\label{spin4} 
\end{eqnarray}
etc...  The classification in terms of spin values is particularly
convenient: (i) under the RG flow the operators with different spin
never mix; (ii) all parameters $h_{m,l}^{a_1...a_l}$ associated
with the components of $P_{m,l}^{a_1...a_l}$ have the same RG
dimension $Y_{m,l}$.  On the other hand, operators with different $m$
but with the same $l$ mix under renormalization.  

The spin-0 operators are already present in the $\Phi^4$ Hamiltonian
(\ref{phi4h}): the RG dimension of $P_{2,0}$ is related to the
correlation length exponent, $Y_{2,0} = y_t=1/\nu$, while the RG
dimension of $P_{4,0}$ (after an appropriate subtraction to cancel the
mixing with $P_{2,0}$) gives the leading scaling correction exponent,
indeed $Y_{4,0}=-\omega$.  The spin-1 perturbation is related to the
external field coupled to the order parameter, thus
$Y_{1,1}=y_h$.~\footnote{ The perturbation $P_{3,1}^a$ is
redundant~\cite{Nicoll-81}, because a Hamiltonian term containing
$P_{3,1}$ can be always eliminated by a redefinition of the field
$\Phi^a$. Anyway, using the equation of motion, one obtains
$Y_{3,1} = (d-2+\eta)/2$.}
Close to four dimensions, thus for small $\epsilon\equiv 4-d$, $Y_{m,l} < 0$
for $l \ge 5$, which implies that the only relevant operators have $l \le 4$.
It is reasonable to assume that this property holds up to $d=3$.  Moreover,
near four dimensions we can use standard power counting to verify that the
perturbation with indices $m,l$ mixes with $P_{m',l}$, $m'\le m$, but their RG
dimensions are significantly smaller. In principle, one should also consider
terms with derivatives of the field, but again one can show that they are all
irrelevant or redundant.

The above arguments show that the most interesting anisotropic
perturbations are represented by the spin-2, spin-3 and spin-4 
operators
\begin{eqnarray}
Q_2^{ab}=P_{2,2}^{ab}, \quad
Q_3^{abc}=P_{3,3}^{abc}, \quad
Q_4^{abcd}=P_{4,4}^{abcd},
\label{ourope} 
\end{eqnarray}
because they provide the leading effects of anisotropy for each spin
sector. As we shall see, the leading RG dimensions within each spin
sector,
\begin{equation}
Y_l\equiv Y_{l,l},
\label{yldef}
\end{equation}
characterize interesting critical behaviors in various physical contexts.
Some $Y_{l}$ have been already estimated by using FT approaches based on
high-order perturbative calculations, and lattice techniques, such as
high-temperature (HT) expansions and Monte Carlo (MC) simulations.  In
Table~\ref{tabrgdimp} we report some results for $N=2,3,4,5$.  In most cases
these results provide already a clear indication of the relevance of the
perturbation, with the only exception of the spin-4 perturbation in the O(3)
universality class, where the value of $Y_4$ is close to zero.  While
high-order FT results indicate the relevance of the spin-4 perturbation, the
MC estimate of $Y_4$ appears compatible with zero.  Since the issue concerning
its relevance is of experimental interest, an accurate determination of
$Y_{4}$ is called for to conclusively settle it.

\begin{table}[tbp]
\caption{Estimates of the RG dimensions $Y_l$ of the couplings $h_{l}$
associated with the leading anisotropic perturbations $Q_l$ for the 
three-dimensional
O($N$) vector universality classes with $N=2,3,4,5$. We report results
obtained by various methods, such as FT perturbative expansions within
$d=3$ and $\epsilon$-expansion schemes, and lattice techniques, such
as high-temperature expansions (HT) and finite-size scaling analyses
of Monte Carlo simulations (FSS MC). Notice that in the MC estimates
of $Y_4$ reported in Ref.~\cite{CH-98} only
 statistical errors are explicitly
given; the authors write that systematic errors are likely of a
similar size.
}
\label{tabrgdimp}
\begin{ruledtabular}
\begin{tabular}{cllll}
\multicolumn{1}{c}{$N$}& \multicolumn{1}{c}{method}&
\multicolumn{1}{c}{$Y_{2}$ (spin 2)}& \multicolumn{1}{c}{$Y_{3}$ (spin 3)}&
\multicolumn{1}{c}{$Y_{4}$ (spin 4)}\\ \colrule 
2 & FT $5^{\rm th}$-order $\epsilon$ expansion &
1.766(6) \cite{CPV-03} & 0.90(2) \cite{DPV-03} & $-$0.114(4)
\cite{CPV-00} \\

  & FT $6^{\rm th}$-order $d=3$ expansion &  
1.766(18) \cite{CPV-03} & 0.897(15) \cite{DPV-03} & $-$0.103(8) 
\cite{CPV-00} \\

  & HT & 1.75(2) \cite{PJF-74} & & \\

  & FSS MC & & & $-$0.171(17) \cite{CH-98} \\

  & FSS MC (this paper) & 1.7639(11) &  0.8915(20) & $-$0.108(6) \\\hline

3 & FT $5^{\rm th}$-order $\epsilon$ expansion &
1.790(3) \cite{CPV-03} & 0.96(3)
\cite{DPV-03} & $\phantom{-}$0.003(4) \cite{CPV-00} \\

  & FT $6^{\rm th}$-order $d=3$ expansion &  
1.80(3) \cite{CPV-03} & 0.97(4) \cite{DPV-03} & 
$\phantom{-}$0.013(6) \cite{CPV-00} \\

  & HT & 1.76(2) \cite{PJF-74} & & \\

  & FSS MC & & & $-$0.0007(29) \cite{CH-98} \\

  & FSS MC (this paper) & 1.7906(3) & 0.9616(10) & 
$\phantom{-}$0.013(4) \\\hline

4 & FT $5^{\rm th}$-order $\epsilon$ expansion &
1.813(6) \cite{CPV-03} & 1.04(5)
\cite{DPV-03} & $\phantom{-}$0.105(6) \cite{CPV-00} \\

  & FT $6^{\rm th}$-order $d=3$ expansion &  
1.82(5) \cite{CPV-03} & 1.03(3) \cite{DPV-03}
& $\phantom{-}$0.111(4) \cite{CPV-00} \\
  & FSS MC & & & $\phantom{-}$0.1299(24) \cite{CH-98} \\

  & FSS MC (this paper) & 1.8145(5) & 1.0232(10) & 
$\phantom{-}$0.125(5) \\ \hline

5 & FT $5^{\rm th}$-order $\epsilon$ expansion &
1.832(8) \cite{CPV-03} & 1.08(4)
\cite{DPV-03} & $\phantom{-}$0.198(11) \cite{CPV-03} \\

  & FT $6^{\rm th}$-order $d=3$ expansion &  
1.83(5) \cite{CPV-03} & 1.07(2)
\cite{DPV-03} & $\phantom{-}$0.189(10) \cite{CPV-03} \\

  & FSS MC & & & $\phantom{-}$0.23(2) \cite{HPV-05} \\

\end{tabular}
\end{ruledtabular}
\end{table}

In this paper we present new accurate estimates of the RG dimensions $Y_l$ of
the anisotropic perturbations for $N=2,3,4$. For this purpose we perform
finite-size scaling (FSS) analyses of Monte Carlo (MC) simulations of lattice
O($N$) spin systems.  We achieve a significant improvement of the accuracy of
the estimates of $Y_l$, essentially by combining the FSS method of
Ref.~\cite{CH-98} with the use of improved Hamiltonians~\cite{ChFiNi}, which
are characterized by the fact that the leading correction to scaling is
suppressed in the asymptotic expansion of any observable near the critical
point.  Our results are also reported in Table~\ref{tabrgdimp}. As we shall
explain later, the errors in the estimates of $Y_l$, and in particular of
$Y_4$, are quite prudential, they are largely dominated by the systematic
error arising from the necessary truncation of the 
Wegner expansions~\cite{Wegner-76} which
provide the asymptotic FSS behavior of the quantities considered.  The results
are a good agreement with the estimates obtained by the analyses of high-order
FT perturbative expansions, in particular with those obtained by resumming
$6^{\rm th}$-order $d=3$ expansions.  Our results show that
spin-4 perturbations in three-dimensional Heisenberg systems are relevant, 
with a quite small
RG dimension $Y_4=0.013(4)$, which may give rise to very slow crossover
effects in systems with small spin-4 anisotropy.  The apparent discrepancy
with the MC result of Ref.~\cite{CH-98}, obtained using the standard
nearest-neighbor O(3) spin model, can be explained by the presence of sizable
scaling corrections. We overcome this problem by using improved lattice
Hamiltonians. The relevance of the spin-4 perturbations is important for
systems with cubic perturbations~\cite{Aharony-76}, and also systems whose
phase diagram presents two transition lines, XY and Ising transition lines,
meeting at a multicritical point~\cite{FN-74}. We shall further discuss these
physical applications later.

The remainder of the paper is organized as follows.  In Sec.~\ref{sec2} we
present the lattice $\phi^4$ spin model which we consider in our MC
simulations, and provide the definitions of the quantities that we
consider in our FSS analyses, in particular, those related to the
spin-$l$ anisotropies.  In Sec.~\ref{sec3} we describe our FSS
analyses of MC simulations which lead to our final estimates already
reported in Table~\ref{tabrgdimp}.  Finally, in the conclusive
Sec.~\ref{discussion} we discuss a number of physical applications of
our results. App. \ref{MCsim} and \ref{fures} contain some details of
the MC simulations, and further results on the critical behavior of
O($N$) vector models.

\section{The lattice model and the estimators of the anisotropy RG
  dimensions}
\label{sec2}

\subsection{Improved lattice O($N$)-symmetric $\phi^4$ models}
\label{models}

In this numerical study of O($N$) vector models with $N=2,3,4$, 
we consider the $\phi^4$ O($N$)-symmetric lattice Hamiltonian
\begin{equation}
{\cal H}_{\phi^4} =
 - \beta\sum_{\left<xy\right>} {\phi}_x\cdot{\phi}_y +
   \sum_x \left[ {\phi}_x^{\,2} +
   \lambda ({\phi}_x^{\,2} - 1)^2\right],
\label{phi4Hamiltonian}
\end{equation}
where $\phi_x$ is an $N$-component real variable, $x$ and $y$ denote sites of
the simple-cubic lattice and $\left<xy\right>$ is a pair of nearest-neighbor
sites. In our convention, the Boltzmann factor is given by $\exp(-{\cal
  H}_{\phi^4})$. For $\lambda=0$ we get the Gaussian model, while in the limit
$\lambda \rightarrow \infty$ the O($N$)-symmetric non-linear $\sigma$ model is
recovered.  For any $0 < \lambda \le \infty$ the model undergoes a continuous
phase transition in the universality class of the O($N$)-symmetric vector
model.

In our FSS analyses we consider cubic $L^3$ lattices with periodic boundary
conditions.  We consider standard finite-volume quantities such as the
magnetic susceptibility and second-moment correlation length related to the
two-point function $G(x-y)\equiv \langle \phi_x \cdot \phi_y\rangle$, i.e
\begin{equation}
\chi  \equiv  \frac{1}{L^3} \langle M^2 \rangle,
\qquad M = \sum_x \phi_x,
\end{equation}
and
\begin{equation}
\xi  \equiv  \sqrt{\frac{\chi/F-1}{4 \sin^2 \pi/L}},
\qquad
F  \equiv  \frac{1}{L^3} \, \biggl\langle
\Big|\sum_x \exp\left(i \frac{2 \pi x_1}{L} \right)
        \phi_x \Big|^2
\biggr\rangle \;\;.
\label{xidef}
\end{equation}
Another standard quantity for FSS analyses is the quartic Binder
cumulant
\begin{equation}
U_{4} \equiv \frac{\langle(M^2)^2\rangle}{\langle M^2\rangle^2}.
\end{equation}
The ratio $\xi/L$ and $U_4$ are RG-invariant phenomenological
couplings, thus their large-volume limit at $T_c$ is universal.  
We also consider quantities defined keeping one of
the phenomenological coupling fixed, in particular keeping the ratio
$\xi/L$ fixed, see, e.g., Ref.~\cite{CHPRV-01}.  We define $\bar{U}_4$
as the Binder cumulant at fixed $\xi/L$.~\footnote{ In previous
studies, see Refs.~\cite{CHPRV-01,CHPRV-02,CHPV-06}, another
RG-invariant quantity turned out to be very useful, i.e. the ratio
$Z_a/Z_p$ of partition functions of a system with anti-periodic
boundary conditions in one direction and periodic ones in the other
two directions and a system with periodic boundary conditions in all
directions.  Since here we focus on the anisotropy, we have not
implemented it to keep the project manageable.}

Improved Hamiltonians are characterized by the fact that the leading
correction to scaling is eliminated in any quantity near the critical point.
Therefore in a MC study, the asymptotic behavior at the phase transition can
be determined more precisely. Improved Hamiltonians were first discussed in
Refs.~\cite{ChFiNi} at the example of the three-dimensional Ising universality
class using high-temperature series expansions.  This idea was first
implemented in MC simulations of $\phi^4$ O($N$)-symmetric lattice models for
$N=2$, $3$ and $4$ in Refs.~\cite{HT-99,H-01}.  In the case of the $\phi^4$
lattice model~(\ref{phi4Hamiltonian}), the improved model is obtained by
tuning the parameter $\lambda$ to the particular value $\lambda^*$, where the
leading $O(L^{-\omega})$ scaling corrections vanish in the FSS behavior of any
quantity. For this purpose, the RG-invariant phenomenological couplings turn
out to be particularly useful.  Indeed, along the critical line
$\beta_c(\lambda)$ or keeping another phenomenological coupling constant, they
behave as
\begin{equation}
R(L,\lambda) = R^* + c(\lambda) L^{-\omega} + ... 
\label{rll}
\end{equation} 
where $c(\lambda)$ is a smooth function of $\lambda$.  Therefore, the equation
$c(\lambda^*)=0$ determines $\lambda^*$.  

The best estimate of $\lambda^*$ for $N=2$ is $\lambda^*=2.15(5)$ obtained in
Ref.~\cite{CHPV-06}.  In the case of $N=3,4$, the MC simulations 
performed for this numerical work lead to a revision of the earlier estimates
of $\lambda^*$, see App.~\ref{fures} for details.  We obtain
$\lambda^*=5.2(4)$ for $N=3$ and $\lambda^*= 20^{+15}_{-6}$ for $N=4$, which
update earlier estimates, respectively $\lambda^*=4.6(4)$ of
Ref.~\cite{CHPRV-02} and $\lambda^*= 12.5(4.0)$ of Ref.~\cite{H-01}.

\subsection{Anisotropy estimators}
\label{anest}

In order to compute the spin-$l$ RG dimensions $Y_l$, we consider appropriate
anisotropy correlators.  We use the magnetization $M^a=\sum_x \phi_x^a$ and
the normalized magnetization $m^a$ defined as
\begin{equation}
m^a\equiv {M^a\over |M|}, 
\label{madef}
\end{equation}
to construct objects with given spin properties, such as
$Q_2^{ab}(m)$, $Q_3^{abc}(m)$, and
$Q_4^{abcd}(m)$, obtained by
replacing $\Phi^a$ with $m^a$ in the expressions of $Q_l$,
cf. Eqs.~(\ref{spin2}), (\ref{spin3}), and (\ref{spin4}).  Then we
consider the correlators
\begin{eqnarray}
&&C_2=\sum_{ab} \left \langle \sum_x  
Q_{2}^{ab}(\phi_x) Q_2^{ab}(m) \right \rangle,
\label{c2def}\\
&&C_3=\sum_{abc} \left \langle \sum_x  Q_{3}^{abc}(\phi_x) Q_3^{abc}(m) 
\right \rangle,
\label{c3def}\\
&&C_4=\sum_{abcd} \left \langle \sum_x Q_{4}^{abcd}(\phi_x)
Q_4^{abcd}(m) \right \rangle,
\label{c4def}
\end{eqnarray}
where $Q_{l}(\phi_x)$ are the operators (\ref{spin2}), (\ref{spin3}),
and (\ref{spin4}) constructed using the lattice variable $\phi^a_x$.
Note that they can be rewritten in term of the angle $\alpha_x$
defined as $\phi_x \cdot m = |\phi_x| {\rm cos} \,\alpha_x$, as
\begin{eqnarray}
&&C_2 =  \left \langle \sum_x |\phi_x|^2 \left( \cos^2\alpha_x - {1\over
  N}\right) \right \rangle, \nonumber \\
&&C_3 = \left \langle \sum_x |\phi_x|^3 \left( \cos^3\alpha_x - {3\over
  N+2}\cos\alpha_x \right) \right \rangle,  \nonumber \\
&&C_4 = \left \langle \sum_x |\phi_x|^4 \left( \cos^4\alpha_x - {6\over
  N+4}\cos^2\alpha_x + {3\over (N+2)(N+4)} \right) \right \rangle. \nonumber
\end{eqnarray}
This expression of $C_4$ shows that it is equal to the improved
quantity considered in Ref.~\cite{CH-98} to compute the RG dimension
of the cubic-symmetric perturbation, apart from a constant factor.
The asymptotic power-law FSS behavior of $C_l$ at $T_c$, i.e.
\begin{eqnarray}
C_l \sim L^{Y_l}, 
\label{clcor}
\end{eqnarray}
allows us to estimate the RG dimension $Y_l$ of the anisotropy associated with
$Q_l$.  Alternative estimators analogous to $C_l$ are also
\begin{equation}
D_l = \sum_{ab...}
\frac{\left \langle \sum_x Q_{l}^{ab...}(\phi_x) Q_l^{ab...}(M) 
\right \rangle}
     {\left \langle M^2 \right  \rangle^{l/2}},\qquad D_l \sim L^{Y_l}.
\label{clcorr2}
\end{equation}
Note that $Q_l^{ab..}(m)$ and $\langle Q_l^{ab..}(M)
\rangle/\langle M^2 \rangle^{l/2}$ are by
construction RG-invariant quantities (with special symmetry properties). Their
derivatives with respect to $h_p$, cf. Eq.~(\ref{freeen}), provide the
correlators $C_l$ and $D_l$.
We also consider the corresponding quantities, $\bar{C}_l$ and
$\bar{D}_l$, at a fixed value of $\xi/L$.

\section{FSS analyses of the anisotropy correlators}
\label{sec3}

In this section we present FSS analyses of high-statistics MC
simulations for the O(2), O(3) and O(4) $\phi^4$ lattice models
(\ref{phi4Hamiltonian}), for values of the parameter $\lambda$
close to $\lambda^*$ providing the suppression of the leading
scaling correction.  App.~\ref{mcalgo} presents some details
of the MC algorithm used in the simulations;
App.~\ref{statsMC} reports the values of the parameters considered in
our MC simulations, the lattice sizes, and the statistics; finally in
App.~\ref{varobs} we discuss the behavior of the variance of the
observables considered, which influenced the strategy of our FSS
analyses of MC simulations.

Most simulations were performed for the O(3) case, where the spin-4 RG
dimension $Y_4$ is close to zero, and therefore high accuracy is needed
to determine its sign.  This task is made particularly hard
by the rapid increase of the cost to get accurate data for $C_4$ and
$D_4$ with increasing the lattice size, essentially due to a
significant increase of their variance, see the discussion in
App.~\ref{varobs}.  As a consequence, our FSS analyses to determine
$Y_4$ are limited to relatively small lattice sizes.  On the other
hand, the systematic error due to the necessary truncation of the
Wegner expansion~\cite{Wegner-76}, see Eq.~(\ref{general}) below, 
of the quantities considered turns out to be significant, and
its reduction requires accurate results for large lattice sizes. This
represents the major limitation for the accuracy of our numerical
determination of $Y_4$.

App. \ref{fures} reports further FSS analyses of the MC simulations which
allow us to update some of the results concerning the O($N$) vector models,
such as the estimates of $\lambda^*$, of the critical exponents and other
universal quantities.

\subsection{General strategy of the FSS analysis}
\label{strategy}

In order to obtain accurate estimates of the universal quantities,
such as the critical exponents and RG dimensions $Y_l$, it is
important to have a robust control of the corrections to the
asymptotic behaviors, which are suppressed by powers of the 
lattice size $L$. 
The behavior of general quantities
introduced to estimate  critical exponents, such as $C_l$ and $D_l$
defined in the previous section, can be expressed by an asymptotic
Wegner expansion~\cite{Wegner-76} as
\begin{equation}
 A(\lambda;L) = c(\lambda) L^{y} [ 1 + a(\lambda)
L^{-\omega} + \sum_{i=2} a_i(\lambda) L^{-\omega_i} ]
\label{general}
\end{equation}
where $y$ is the leading universal exponent that one wants to
accurately estimate.  In the case of O(2), O(3) and O(4) vector models
the leading scaling correction exponent is given by $\omega\approx
0.8$, see Table~\ref{tabcexp}. Numerical approaches based on improved
Hamiltonians allow us to suppress these leading scaling corrections,
and also those related to $n\omega$, where $n=2,3,4,...$,
whose coefficients behave as $(\lambda-\lambda^*)^n$.
The next-to-leading correction is
controlled by the exponent $\omega_2$, estimated in Ref.~\cite{NR-84}
by $\omega_2\approx 1.8$, see Table~\ref{tabcexp}. Then there are well
established corrections with $\omega_i \approx 2$, for example 
related to the breaking of spatial rotational invariance in cubic
lattice systems~\cite{CPRV-99}, but also to analytic backgrounds,
etc...  Moreover, in the case of the spin-$l$ anisotropy correlators,
we may also have scaling corrections induced by higher-dimensional
spin-$l$ operators, such as $P_{l+2,l}$, cf. Eq.~(\ref{pmldef}).  On
the basis of a dimensional analysis around four dimensions, 
they are expected to
give rise to scaling corrections suppressed by powers
$\kappa_l=2+O(\epsilon)$, as also shown by the $O(\epsilon)$
calculation of the difference of the RG dimensions of the anisotropy
operators $P_{l+2,l}$ and $P_{l,l}$, which is~\footnote{We note that
within $\epsilon$ expansion the operator $P_{l+2,l}$ mixes with other
spin-$l$ operators containing derivatives (two derivatives instead of
$\Phi^2$), but this mixing contributes to $O(\epsilon^2)$.}
\begin{equation}
Y_{l+2,l} - Y_{l,l} = -2 -
\epsilon 6(l-1)/(N+8) + O(\epsilon^2).
\label{dyll}
\end{equation}
In known cases for the spin-0,1,2
sectors, the difference between RG dimensions of the same sector remains close
to their four dimensional values.  
Therefore, as a prudential procedure, after curing the
residual $O(L^{-\omega})$ scaling corrections, see also below, we must
consider possible $O(L^{-\kappa})$ scaling corrections with $\kappa\gtrsim
1.6$.

\subsubsection{Residual leading scaling corrections in approximately
improved Hamiltonians}
\label{rescorr}

Residual leading scaling corrections are generally present due to the
fact that $\lambda^*$ is only known approximately, and also because
the MC simulations are usually performed close but not exactly at the
best estimate of $\lambda^*$, which is usually determined at the end
of the MC simulations. For example, in the case $N=3$ our best
estimate is $\lambda^*=5.2(4)$, while most MC simulations were
performed at $\lambda=4.5$, and others at $\lambda=4$ and $\lambda=5$
for smaller lattices to determine $\lambda^*$.

The residual $O(L^{-\omega})$ corrections, due to the fact that $\lambda$ is
close but does not coincide with its optimal value $\lambda^*$, can be further
suppressed as follows.  The basic idea is that leading corrections to scaling
can be best detected by analyzing the Binder cumulant $\bar{U}_4$ at a fixed
value of $\xi/L$.  At a generic $\lambda=\lambda_0$ we have
\begin{equation}
\bar{U}_4(\lambda_0;L) = \bar{U}_4^* + a_U(\lambda_0) L^{-\omega} + ...,
\label{ubar4}
\end{equation}
where $\bar{U}_4^*$ is the universal large-volume limit on a periodic
$L^3$ box at fixed $\xi/L$, which of course depends on which value of
$\xi/L$ is chosen.  Then, we consider a pair $\lambda_1$, $\lambda_2$, where
one of the two values may be equal to $\lambda_0$, and the differences
\begin{equation}
\Delta_U(\lambda_1,\lambda_2;L)= \bar{U}_4(\lambda_2;L)-\bar{U}_4(\lambda_1;L)
\end{equation}
where the leading large-volume contributions cancel, thus they behave
as
\begin{equation}
\Delta_U(\lambda_1,\lambda_2;L) = b_U(\lambda_1,\lambda_2) L^{-\omega} + ... \; .
\label{leaddiff}
\end{equation}
The amplitude
$b_U(\lambda_1,\lambda_2)=a_U(\lambda_2)-a_U(\lambda_1)$
can be estimated by fitting the data to  (\ref{leaddiff}).
Finally, we take ratios 
\begin{equation}
r_A(\lambda_1,\lambda_2;L) = {A(\lambda_2;L)\over A(\lambda_1;L)}
\end{equation}
of the quantity $A$ that we intend to correct
to eliminate the residual $O(L^{-\omega})$ corrections.
Their data can be fitted to its large-$L$ behavior
\begin{equation}
 r_A(\lambda_1,\lambda_2;L) = {c(\lambda_2)\over c(\lambda_1)} 
\left[ 1 + b(\lambda_1,\lambda_2) L^{-\omega}\right],
\end{equation}
where $b(\lambda_1,\lambda_2) = a(\lambda_2)-a(\lambda_1)$ and $a(\lambda)$ is
the amplitude of the $O(L^{-\omega})$ corrections, cf. Eq.~(\ref{general}).
Notice that it is simpler to extract $b(\lambda_1,\lambda_2)$ than
$a(\lambda)$ from the numerical data, because, beside the cancellation of the
power divergence $L^y$, also subleading corrections cancel to a large extent.
Now we use the universality of ratios of correction amplitudes, which implies
\begin{equation}
\frac{a(\lambda_0)}{a_U(\lambda_0)} = 
\frac{b(\lambda_1,\lambda_2)}{b_U(\lambda_1,\lambda_2)}.
\end{equation}
In order to eliminate the leading 
$O(L^{-\omega})$ corrections from $A$, we construct
\begin{equation}
\label{tildecorrect}
{\cal I}_A(\lambda_0;L) = A(\lambda_0;L) 
\left[1 - \frac{b(\lambda_1,\lambda_2)}{b_U(\lambda_1,\lambda_2)} 
a_U(\lambda_0)L^{-\omega}\right]
\end{equation}
This procedure eliminates the leading $O(L^{-\omega})$ scaling
corrections, allowing us to neglect them in the fits of the data of 
${\cal I}_A(\lambda_0;L)$ 
to estimate the leading exponent $y$.~\footnote{The coefficient
  $c \equiv a_U(\lambda_0) b(\lambda_1,\lambda_2)/b_U(\lambda_1,\lambda_2)$ is
  numerically determined with an error $\Delta c$, which is usually dominated
  by the uncertainty on $a_U(\lambda_0)$. This error can be taken into account
  by computing ${\cal I}_A(\lambda_0;L)$ using $c$ and $c\pm \Delta c$.  The
  difference between the results of their fits is essentially related to the
  error due to the uncertainty of our estimate for $\lambda^*$, since also the
  uncertainty of the estimate of $\lambda^*$ is mainly caused by the error of
  $a_U(\lambda_0)$, see also App.~\ref{lambda*}.}

We also mention that alternative procedures, based on the idea of defining
improved observables with suppressed leading scaling corrections, are outlined
in Refs.~\cite{CHPV-06,HPPV-07}.

\subsubsection{Next-to-leading corrections}
\label{nextrescorr}

Next-to-leading corrections arise from the
term associated with $\omega_2\approx 1.8$, and the others with exponents close to
two.  In the fits of the data, even with high statistics data as we have here,
only
a very limited number of correction terms can be taken into account.  The
truncation of Eq.~(\ref{general}) leads to systematic errors in the results
for the exponent $y$.

One way to control these systematic errors is to study several quantities
$A^{(n)}$ that have the same critical behavior:
\begin{equation}
\label{general2}
 A^{(n)}(L) = c_n L^{y} (1 + \sum_i a_{ni} L^{-\omega_i})
\end{equation}
In general one might expect that for different $A^{(n)}$ the coefficients
$a_{ni}$ are different. Therefore the variation of the estimate for $y$
obtained by fitting several $A^{(n)}$ provides an estimate of the systematic
error.  However, in our case we have only the two quantities $C_l$ and $D_l$,
which are closely related. Therefore we would like to estimate the systematic
error by fitting a single quantity.  To this end we consider the Ansatz
\begin{equation}
\label{assume}
 A(L) = c L^{y} (1 + a L^{-\omega} + a_{2,{\rm eff}}
 L^{-\omega_{2,{\rm eff}}})
\end{equation}
(for improved models $a=0$), with
\begin{equation}
\omega_{2,{\rm eff}} \ge 1.6
\label{omega2eff}
\end{equation}
Barring an unlike significant cancellation between
different correction terms, there must be a value of $\omega_{2,{\rm
    eff}}>1.6$ such that $y$ takes its correct value.  Since we expect that,
as long as correction are small, the resulting $y$ is a monotonic function of
$\omega_{2,{\rm eff}}$, we use the results obtained for $\omega_{2,{\rm
    eff}}=1.6$ and $\omega_{2,{\rm eff}}=\infty$ (i.e. without the term
$c_{2,{\rm eff}} 
L^{-\omega_{2,{\rm eff}}}$) as bounds for the correct result for
$y$.

\subsection{Results for the spin-$l$ RG dimensions}
\label{respinl}

\subsubsection{The O(3) model}
\label{o3mres}

\begin{figure}[tbp]
\includegraphics*[scale=\graphicscale]{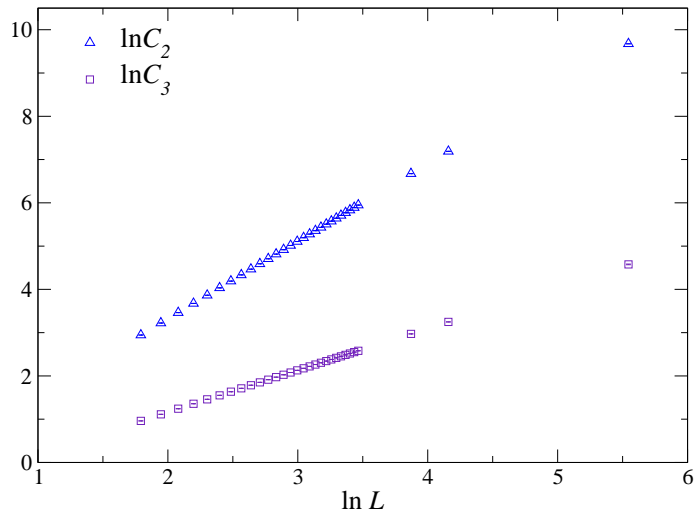}
\caption{(Color online)
Log-Log plots of $C_2$ and $C_3$ 
versus $L$ at $\beta_c$ and $\lambda=4.5$.
The errors of the data are hardly visible.
}
\label{dataC23}
\end{figure}

\begin{figure}[tbp]
\includegraphics*[scale=\graphicscale]{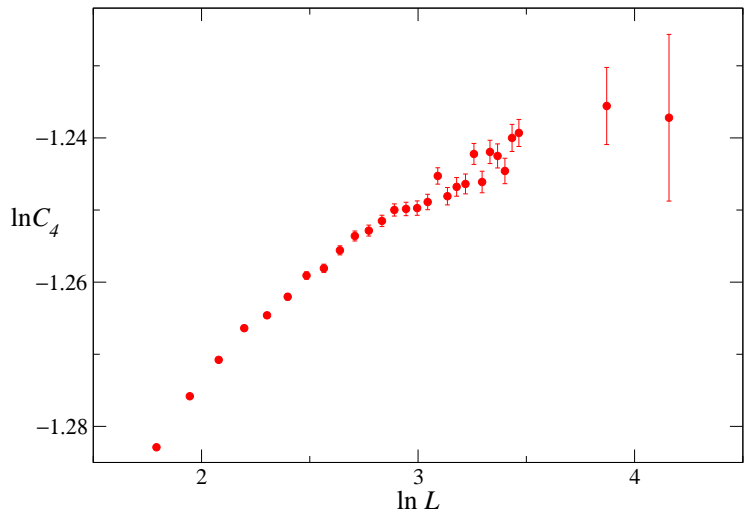}
\caption{(Color online)
Log-Log plots of $C_4$ 
versus $L$ at $\beta_c$ and $\lambda=4.5$
}
\label{dataC4}
\end{figure}

To begin with we present the FSS analysis of the data for the O(3)
model.  In order to give an idea of the quality of our data, we show
the data of $C_l$ at $\beta_c$ and $\lambda=4.5$ in
Figs.~\ref{dataC23} and \ref{dataC4}.  In all cases, including $C_4$,
the data clearly increase with increasing $L$, indicating the
relevance of the perturbation.  Note that the error of
$C_4$ is rapidly increasing with increasing $L$,
see App.~\ref{varobs} for details.

We analyze various quantities to estimate the RG dimensions $Y_l$: the
original quantities $C_l$ and $D_l$ introduced in Sec.~\ref{anest},
their counterpart $\bar{C}_l$ and $\bar{D}_l$ computed at the fixed
value $\xi/L=0.5644$ (which is a good estimate of the large-volume
limit of $\xi/L$ at $\beta_c$, see App.~\ref{fures}), and also the
quantities $C_{l,{\rm imp}} = \bar{U}_4^x \bar{C}_l$ and $D_{l,{\rm
imp}} = \bar{U}_4^x \bar{D}_l$ again taken at $\xi/L=0.5644$ where the
exponent $x$ is chosen to further suppress the leading corrections
(see Refs.~\cite{CHPV-06,HPPV-07} for details).  In principle, the
latter quantities should be more suitable for the numerical
analysis. Indeed, by fixing $\xi/L=0.5644$ we avoid the error due to
the uncertainty of $\beta_c$, and by the construction of the improved
observables the effect of the uncertainty of $\lambda^*$ is strongly
reduced.  However also subleading corrections vary, and,
unfortunately, they become numerically larger in these cases.
Nethertheless it is useful to study these quantities.  Since the
amplitudes of corrections change, these modified quantities give us
additional control over the systematic error that is caused by
truncated ansaetze.

Let us now discuss the analysis of the quantities $D_l$ in some detail.
In the case of the quantities $C_l$ we proceed in a similar way.
Following the discussion of section \ref{rescorr} we first analyze
the ratios 
\begin{equation}
 r_{D_l} =  \frac{D_l(\lambda=5,\beta=0.687564)}
           {D_l(\lambda=4,\beta=0.68439)}
\end{equation}
where $\beta=0.687564$ and $\beta=0.68439$ are the estimates for 
$\beta_c$ given in Table V of Ref.~\cite{CHPRV-02} and
\begin{equation}
 r_{\bar{D}_l} =  \frac{\bar D_l(\lambda=5)}
           {\bar D_l(\lambda=4)} \;\;. 
\end{equation}
We fit these ratios to the ansatz
\begin{equation}
 r = c (1 + b L^{-\omega}) 
\end{equation}
where we set $\omega=0.79$.  To give an idea how accurately the
coefficient $b$ can be determined let us discuss a few examples.  In
the case of $D_2$ a fit of all data with $L\ge L_{min}$ with
$L_{min}=6$ gives the result $b=-0.00841(38)$ and
$\chi^2/$DOF$=3.80/9$. Increasing $L_{min}$ the estimate of $b$ changes
very little, for example, for $L_{min}=8$ we obtain $b=-0.00830(64)$ and
$\chi^2/$DOF$=3.58/7$.  In the following analysis we shall
assume $b=-0.0083(7)$.  In the case of $\bar{D}_2$ we obtain for
$L_{min}=7$ the result $b=0.00348(24)$ and $\chi^2/$DOF$= 8.48/8$ and
for $L_{min}=9$ the results $b=0.00407(41)$ and $\chi^2/$DOF$=
2.34/6$.  In the following analysis we shall assume $b=0.004(1)$.  It
is interesting to observe that, by taking $D_2$ at $\xi/L=0.5644$
instead of $\beta_c$, even the sign of the correction amplitude changes.
For $D_4$ we obtain $b=0.0313(45)$ and $\chi^2/$DOF$= 7.49/8$ using
$L_{min}=7$. The result changes little when we increase $L_{min}$. For
example, we get $b=0.0303(93)$ and $\chi^2/$DOF$= 7.42/6$ for
$L_{min}=9$. In the following we shall assume $b=0.03(1)$.  For $\bar
D_4$ we get instead $b=0.05(1)$.  Note that also 
the correction amplitudes of $D_4$ and $\bar D_4$
are different.

In order to compute the quantities ${\cal I}_{D_l}$ and ${\cal
I}_{\bar{D}_l}$, defined as in Eq.~(\ref{tildecorrect}) to suppress the
residual leading scaling corrections, we use $b_U(5,4)=-0.01126(4)$
and $a_U(4.5)=0.007(4)$ as obtained in appendix \ref{lambda*}. In the
product $\bar U_4^x \bar D_l$ the choice $x = - \bar U_4^* b/b_U$
eliminates leading corrections to scaling. The advantage of this
quantity is that it does not require $a_U$, which is affected by a
relatively large error.

Next we have fitted the resulting quantities with the ansaetze 
\begin{equation}
\label{fitD1}
{\cal I}_{D_l}(\lambda_0;L) \equiv D_l(\lambda_0;L) 
\left[1 - \frac{b_l(\lambda_1,\lambda_2)}{b_U(\lambda_1,\lambda_2)} 
a_U(\lambda_0)L^{-\omega}\right] 
= a L^{Y_l} 
\end{equation}
and
\begin{equation}
\label{fitD2}
{\cal I}_{D_l} = a L^{Y_l} (1 + d L^{-1.6}) 
\end{equation}
and correspondingly for the quantities ${\cal I}_{\bar{D}_l}$ and $\bar
U_4^x \bar D_l$. The effect of the uncertainties of $\beta_c$, and the
quantities $a_U$, $b_U$, $b_l$ need to construct ${\cal I}_{D_l}$, 
${\cal I}_{\bar{D}_l}$ and $\bar U_4^x \bar D_l$, are
estimated by varying these input parameters. E.g. in order to
estimate the uncertainty of ${\cal I}_{\bar{D}_l}$ induced 
by the uncertainty of $a_U$, we have repeated the fits using data
where we have used in eq.~(\ref{tildecorrect}) the central 
value of $a_U$ plus its error instead of the central value.

In Table \ref{tabD2} we report results of fits for ${\cal I}_{D_2}$, 
${\cal I}_{\bar{D}_2}$, and $\bar U_4^x \bar D_2$.  We note that the estimates of
$Y_2$ obtained by the two fits and the three quantities differ by
larger amounts than their statistical errors. Hence systematic errors
are more important than the statistical one.
Taking into account also the results obtained for $C_4$ and the 
quantities derived from it we arrive at our final estimate 
$Y_2 = 1.7906(3)$
which covers most of the acceptable fits and also takes into account
the uncertainties in the construction of ${\cal I}_{D_2}$, ${\cal
I}_{\bar{D}_2}$ and $\bar U_4^x \bar D_2$.  In a similar way we arrive at the
estimate $Y_3=0.9616(10)$ of the spin-3 RG dimension.

\begin{table}
\caption{Fits of  ${\cal I}_{D_2}$ (column 2 and 3), ${\cal I}_{\bar{D}_2}$ 
(column 4 and 5) and $\bar U_4^x \bar D_2$ (column 6 and 7)
with the ansaetze~(\ref{fitD1}) and (\ref{fitD2}). We give the $L_{min}$ of 
the fit, which is typically the smallest $L_{min}$ that produces an
acceptable fit and the result for $Y_2$. 
}
\label{tabD2}
\begin{ruledtabular}
\begin{tabular}{ccccccc}
 ansatz&$L_{min}$ &  $Y_2$  &    $L_{min}$ &  $Y_2$  &  $L_{min}$ &  $Y_2$   \\
\hline
(\ref{fitD1})  & 24 & 1.79067(5) &  28 & 1.79078(3) &32 & 1.79080(5) \\
(\ref{fitD2})  & 12 & 1.79019(7) &   8 & 1.79053(2) & 8 & 1.79049(2) \\
\end{tabular}
\end{ruledtabular}
\end{table}

\begin{table}
\caption{Fits of  ${\cal I}_{D_4}$ (column 2 and 3), ${\cal I}_{\bar{D}_4}$ 
(column 4 and 5) and $\bar U_4^x \bar D_4$ (column 6 and 7)
with the ansaetze~(\ref{fitD1}) and (\ref{fitD2}). We give the $L_{min}$ of
the fit, which is typically the smallest $L_{min}$ that produces an
acceptable fit and the result for $Y_4$.
}
\label{tabD4}
\begin{ruledtabular}
\begin{tabular}{ccccccc}
 ansatz&$L_{min}$ &  $Y_4$  &    $L_{min}$ &  $Y_4$  &  $L_{min}$ &  $Y_4$   \\
\hline
(\ref{fitD1})  &14 &  0.0143(8) & 14  & 0.0142(8)   & 16 & 0.0160(10)  \\
(\ref{fitD2})  &12 &  0.0122(26)& 12  & 0.0127(25)  & 12 & 0.0122(26) \\
\end{tabular}
\end{ruledtabular}
\end{table}

Finally, let us discuss the analysis leading to our estimate of $Y_4$.  In
Table \ref{tabD4} we give some results of the fits with the ansaetze~(\ref{fitD1})
and (\ref{fitD2}).  As our final result we quote $Y_4=0.013(4)$ which covers
all estimates given in Table \ref{tabD4}. The uncertainties in the
construction of ${\cal I}_{D_4}$, ${\cal I}_{\bar{D}_4}$ and $\bar U_4^x \bar
D_4$ are taken into account.  Furthermore, this estimate is fully consistent
with the results obtained from the analysis of ${\cal I}_{C_4}$, ${\cal
  I}_{\bar{C}_4}$ and $\bar U_4^x \bar C_4$.

We conclude with a few remarks on the possibility of further improving the
estimate of $Y_4$.  Its accuracy is essentially limited by the fact that the
variances of the correlators $C_4$ and $D_4$ rapidly increase with increasing
lattice size, not allowing us to get accurate results for large lattices,
indeed extremely high statistics are necessary for $L \gtrsim 32$ already.
Thus, the reduction of the systematic error due to the truncation of the
Wegner expansion appears quite problematic, because it can only get reduced by
accurate results for larger lattice sizes.  One purely technical way in this
direction could be the simulation with local algorithms (Metropolis + many
overrelaxation sweeps) on GPUs (Graphics cards).

\subsubsection{The O(2) and O(4) models}
\label{o24mres}

In the cases of the XY and O(4) universality classes we have determined the
exponents along similar lines, obtaining the results reported in Table
\ref{tabrgdimp}.  We only mention that, since in the case of the XY
universality class, $\lambda^*$ and $\beta_c$ at $\lambda=2.1$ are accurately
known \cite{CHPV-06}, we abstained from analyzing the quantities $\bar{U}_4^x
\bar{C}_l$ and $\bar{U}_4^x \bar{D}_l$.  In the case of the O(4) universality
class the situation is different; here we do not have a very precise estimate
of $\lambda^*$ and also $\beta_c$ is only moderately well known at
$\lambda=12.5$, where most of our simulations are performed.  Therefore we
have based our analysis on $\bar{C}_l$ and $\bar{D}_l$ and the improved
quantities $\bar{U}_4^x \bar{C}_l$ and $\bar{U}_4^x \bar{D}_l$, where the
quantities are taken at $\xi/L=0.547$.

\section{Conclusions and discussion of some applications}
\label{discussion}

In this paper we study the effects of anisotropic perturbations in
three-dimensional O($N$)-symmetric vector models, which cannot be related to an
external vector field coupled to the order parameter, but are
represented by composite operators with more complex transformation
properties under the O($N$) group.  For the models with $N=2,3,4$, we
determine the RG dimensions $Y_l$ of the
anisotropic perturbations associated with the first few spin values of
the representations of the O($N$) group, because the lowest spin
values give rise to the most important effects.  This is the first
numerical study based on MC simulations for the spin-2 and spin-3
perturbations, while MC results for spin-4 operators were already reported
in Ref.~\cite{CH-98}.

We present FSS analyses of MC simulations of improved Hamiltonians
with suppressed leading corrections to scaling, which allows us to
achieve a robust control of the systematic errors arising from scaling
corrections.  Our results are reported in Table~\ref{tabrgdimp},
together with earlier results by various approaches.  They are in good
agreement with the estimates obtained by field-theoretical methods, by
resumming high-order perturbative series.  Our results 
show that spin-4 perturbations in three-dimensional Heisenberg systems
are relevant, with a quite small RG dimension $Y_4=0.013(4)$, which
may give rise to very slow crossover effects in systems with small
spin-4 anisotropy.

In the following we discuss a number of physical systems where the results of
this paper for the anisotropic perturbations can be used to infer the critical
behavior of some physically interesting quantities.

\subsection{Critical exponents of secondary order parameters}
\label{secpar}

Beside the standard critical exponents associated with the order
parameter, density wave XY systems allow to measure the
higher-harmonic critical exponents related to secondary order
parameters, which can be theoretically represented by polynomials of
the order parameter with spin representation higher than one, such as
the spin-$l$ operators $Q_l(\phi_x)$, cf. Eqs.~(\ref{spin2}-\ref{spin4}).

The behavior at zero-momentum of the correlation functions involving
the operators $Q_l(\phi_x)$ can be described by introducing an appropriate
external field $h_l$ coupled with $Q_l(\phi_x)$, and writing the singular
part of the free energy as in Eq.~(\ref{freeen}). 
Then, differentiating with respect to $h_l$, we
obtain the behavior of the secondary magnetizations in the broken
phase,
\begin{eqnarray}
\langle Q_l(\phi_x) \rangle \sim |t|^{\beta_l},\qquad \beta_l = \nu (d - Y_l).
\label{betal}
\end{eqnarray}
Our estimates of the RG dimensions $Y_l$ for the XY universality class,
$Y_2=1.7639(11)$, $Y_3=0.8915(20)$ and $Y_4=-0.108(6)$, 
give 
\begin{eqnarray}
\beta_2= 0.8303(8),\quad 
\beta_3=1.4163(13), \quad
\beta_4=2.09(4). 
\label{betaest}
\end{eqnarray}
Moreover, the nonanalytic scaling behaviors of spin-$l$
susceptibilities are
\begin{eqnarray}
\chi_l \equiv \sum_x \langle Q_l(\phi_0) Q_l(\phi_x) \rangle\sim
|t|^{-\gamma_l},\qquad \gamma_l = \nu (2Y_l-d),
\label{gammal}
\end{eqnarray}
with 
\begin{eqnarray}
\gamma_2=0.3545(15),
\quad 
\gamma_3=-0.817(3),
\quad
\gamma_4=-2.160(8).  
\label{gammaest}
\end{eqnarray}
Note that the power law $|t|^{-\gamma_l}$ in the
susceptibility $\chi_l$ represents the leading term only if
$\gamma_l>0$, otherwise the nonuniversal analytic contributions
provide the dominant behavior, see, e.g., Ref.~\cite{CPV-02}. We also
mention that the structure factor, obtained by Fourier
transforming the correlation function $G_l(x-y) = \langle Q_l(\phi_x) 
Q_l(\phi_y)
\rangle$, is expected to behave as $\widetilde{G}_l(q) \sim
|t|^{-\gamma_l} f_l(q\xi)$, where $f_l$ is a universal function, see
Ref.~\cite{CPV-02} and references therein.

Discussions of the experimental systems and results for the
higher-harmonic exponents can be found in
Refs.~\cite{PV-r,CPV-02,DPV-03}.  The experimental estimates are in
substantial agreement with the theoretical results.  Here we only
mention a few of them.
Analyses~\cite{Brock-etal-86,ABBL-86,Brock-et-al-89} of the
experimental data near the smectic-C-tilted-hexatic-I transition
provided estimates of the crossover exponent $\phi_l=Y_l\nu$. By
replacing $\nu=0.6717$, they give $Y_2=1.7(1)$ and $Y_3=0.6(3)$.
In Ref.~\cite{ZMHCLGGS-96} the estimates $\beta_2=0.87(1)$ and
$\beta_3=1.50(4)$ were obtained for Rb$_2$ZnCl$_4$.

\subsection{Magnets with cubic symmetry}.
\label{cubic}

The magnetic interactions in crystalline solids with cubic symmetry, like iron
or nickel, are usually modeled by using the O(3)-symmetric Heisenberg
Hamiltonian with short-range spin interactions, such as
\begin{equation}
H_{\rm spin} = -J \sum_{\langle ij\rangle} S_i \cdot S_j
\label{heisenberg}
\end{equation}
where $S^2=1$ and the sum is over nearest neighbors.  However, this is a
simplified model, since other interactions are present.  Among them, the
magnetic anisotropy that is induced by the lattice structure (the so-called
crystal field) is particularly relevant experimentally, see, e.g.,
Ref.~\cite{Chikazumi-book}.  In cubic-symmetric lattices it gives rise to
additional single-ion contributions, the simplest one being
\begin{equation}
\sum_i \sum_a S_i^{a\,4}.
\label{addcu}
\end{equation}
These terms are usually not considered when the critical behavior of
cubic magnets is discussed.  However, this is strictly justified only
if these nonrotationally invariant interactions, that have the reduced
symmetry of the lattice, are irrelevant in the RG sense.  The
corresponding cubic-symmetric perturbation $\sum_a \Phi^{a\,4}$ to the
O($N$) theory is a particular combination of spin-4 operators
$P_{4,4}^{abcd}$ and of the spin-0 term $P_{4,0}$,
\begin{equation}
\sum_a \Phi^{a\,4} = 
\sum_{a=1}^N P_{4,4}^{aaaa}(\Phi) + {3\over N+2} P_{4,0}(\Phi) 
\label{cubicper}
\end{equation}
Since $P_{4,0}$ is always irrelevant, the relevance of the
cubic-symmetric anisotropy is related to the value of the spin-4 RG
dimension $Y_{4}$, and in particular to its sign. Our results, and in
particular $Y_{4}=0.013(4)$ for the O(3) universality
class, show that the
cubic perturbation is relevant at the three-dimensional O($N$) 
fixed point when $N\ge
3$, confirming earlier FT results~\cite{CPV-00,PS-00,FHY-00,KS-95}.
This implies that for $N\ge 3$ the asymptotic critical behavior is
described by another cubic-symmetric fixed point, see, e.g.,
Refs.~\cite{PV-r} for a general discussion of the RG flow in the
$\Phi^4$ theories with cubic-symmetric anisotropy.  However,
differences between the Heisenberg and cubic critical exponents are
very small \cite{CPV-03}, for example $\nu$ differs by less than
0.1\%, which is much smaller than the typical experimental error for
Heisenberg systems \cite{PV-r}.  Therefore, distinguishing the cubic
and the Heisenberg universality class is very hard in
experiments.

\subsection{Multicritical phenomena in O($n_1$)$\oplus$O($n_2$)-symmetric
systems}
\label{mcb}

The competition of distinct types of ordering gives rise to
multicritical behaviors.  The multicritical behavior
arising from the competition of two types of ordering characterized by
O($n$) symmetries is determined by the RG flow of the most general
O($n_1$)$\oplus$O($n_2$)-symmetric LGW Hamiltonian involving two
fields $\phi_1$ and $\phi_2$ with $n_1$ and $n_2$ components
respectively, i.e.~\cite{FN-74}
\begin{eqnarray}
{\cal H}_{mc} = &&\int d^d x \Bigl[ 
\case{1}{2} ( \partial_\mu \phi_1)^2  + \case{1}{2} (
\partial_\mu \phi_2)^2 + \case{1}{2} r_1 \phi_1^2  
 + \case{1}{2} r_2 \phi_2^2 
+ u_1 (\phi_1^2)^2 + u_2 (\phi_2^2)^2 + w \phi_1^2\phi_2^2 \Bigr].
\label{bicrHH}
\end{eqnarray}
A multicritical point (MCP) is achieved when $r_1$ and $r_2$ are tuned
to their critical value, and the corresponding multicritical behavior
is determined by the stable FP of the RG flow of the quartic
parameters.  It may occur at the intersection of two critical lines
characterized by different O($n_1$) and O($n_2$) order parameters.

An interesting possibility is that the stable FP has O($n_1+n_2$) symmetry, so
that the symmetry gets effectively enlarged approaching the MCP.  The
stability properties of the O($n_1+n_2$) symmetric FP can be inferred by
noting~\cite{CPV-03} that the Hamiltonian (\ref{bicrHH}) contains combinations
of spin-2 and spin-4 polynomial operators with respect to the O($n_1+n_2$)
group, which are invariant under the symmetry O($n_1$)$\oplus$O($n_2$).
Defining $\Phi$ as the $(n_1+n_2)$-component field $(\phi_1,\phi_2)$, they are
given by the spin-0 operators $\Phi^2$ and $(\Phi^2)^2$, by the spin-2
operators
\begin{eqnarray}
O_{2,2}= 
\sum_{a=1}^{n_1} P_{2,2}^{aa} = 
\phi_1^2-{n_1\over n_1+n_2} \Phi^2,\qquad
O_{4,2}= \Phi^2 O_{2,2} ,\label{spin2mc}
\end{eqnarray}
and by the spin-4 operator
\begin{eqnarray}
O_{4,4}=  \sum_{a=1}^{n_1} \;\sum_{b=n_1+1}^{n_2} P_{4,4}^{aabb} = 
\phi_1^2 \phi_2^2- {\Phi^2 (n_1 \phi_2^2+n_2 \phi_1^2)\over n_1+n_2+4}+
{n_1 n_2 (\Phi^2)^2 \over (n_1+n_2+2)(n_1+n_2+4)}.
\label{spin4mc}
\end{eqnarray}
The O($n_1+n_2$) FP controls the multicritical behavior if it is stable
against the fourth-order perturbations, and, in particular, the dominating
spin-4 perturbation $O_{4,4}$, (the perturbation $O_{4,2}$ is expected to be
irrelevant after the subtraction of its lower-dimension spin-2
content~\cite{CPV-03}).

Our FSS MC results for the spin-4 RG dimensions $Y_4$ (see
Table~\ref{tabrgdimp}), and, in particular, that for the O(3)
universality class, provide a conclusive evidence that $Y_4>0$ for
$n_1+n_2\ge 3$, confirming earlier indications from FT
computations~\cite{CPV-03}.  Therefore the enlargement of the symmetry
O($n_1$)$\oplus$O($n_2$) to O($n_1+n_2$) does not occur, unless an
additional parameter is tuned beside those associated with the
quadratic perturbations.  We may observe an enlargement of the
symmetry to $O(2)$ only when two Ising lines meet.  In this case the
RG dimension $Y_2$ of the spin-2 operator $O_{2,2}$ provides the
crossover exponent $\phi=\nu Y_{2}=1.1848(8)$ at the MCP.

These results can be applied to the study of the phase diagram of anisotropic
antiferromagnets in a uniform magnetic field $H_\parallel$ parallel to the
anisotropy axis, which present a MCP in the $T-H_\parallel$ phase diagram,
where two critical lines belonging to the XY and Ising universality classes
meet \cite{FN-74,NKF-74}.  Experimental realizations of these systems are
reported in Refs.~\cite{RG-77,KR-79,OPSB-78}, which typically show phase
diagrams with a bicritical MCP. The initial hypothesis of an enlarged O(3)
symmetry at the MCP, on the basis of low-order FT calculations~\cite{NKF-74},
was then questioned by high-order FT computations~\cite{CPV-03} (see also
Ref.~\cite{FHM-08}), indicating a very weak instability of the O(3) FP.  This
instability was then questioned by the numerical MC study of
Ref.~\cite{Selke-11}, where evidence of a O(3)-symmetric bicritical point is
claimed in the phase diagram of the so-called XXZ model, which models
anisotropic antiferromagnets in an external field, showing a MCP where an XY
and an Ising transition line meet.  Actually, this result was one of the major
motivation of this numerical work to further check the relevance of the spin-4
perturbation at the O(3) FP, because an asymptotic O(3) multicritical behavior
requires $Y_4<0$.  Our MC results fully confirm earlier high-order FT results,
i.e. the relevance of the spin-4 O(3)-breaking term which are generally
present in these models.  This implies that a bicritical point in the
Heisenberg universality class is excluded, unless one achieves a complete
cancellation of the spin-4 term by an appropriate fine tuning.

As inferred by FT calculations, the actual stable FP has a biconical
structure~\cite{CPV-03}.  A quantitative analysis of the biconical FP
shows that its critical exponents are very close to the Heisenberg
ones.  For instance, the correlation-length exponent $\nu$ differs by
less than 0.001 in the two cases.  Thus, it should be very hard to
distinguish the biconical from the O(3) critical behavior in experiments
or numerical works based on Monte Carlo simulations.

The crossover exponent describing the crossover from the unstable O(3)
critical behavior is very small, i.e. $\phi_{4}=\nu Y_{4} = 0.009(3)$,
so that systems with a small effective breaking of the O(3) symmetry
show a very slow crossover towards the biconical critical behavior or,
if the system is outside the attraction domain of the biconical FP,
towards a first-order transition. Thus, they may show the eventual
asymptotic behavior only for very small values of the reduced
temperature.  Likely, the numerical analysis of Ref.~\cite{Selke-11}
was just observing crossover effects.

\section{Acknowledgements}
This work was supported by the DFG under the grant No HA 3150/2-2.

\appendix

\section{Monte Carlo simulations}
\label{MCsim}

\subsection{Monte Carlo algorithm}
\label{mcalgo}

As Monte Carlo algorithm we use a hybrid of the local Metropolis,
the local overrelaxation and the single cluster \cite{Wolff} algorithm. The
proposals for the local Metropolis update are given by
\begin{equation}
\phi_x' = \phi_x + s r_x
\end{equation}
where $s$ controls the step size and the components of the random vector $r_x$ 
are uniformly distributed in the interval  $[-0.5,0.5]$.
This proposal is accepted with the standard acceptance probability
\begin{equation}
 P_{acc} = \mbox{min}[1,\exp(-\Delta \cal H)] \;\;. 
\end{equation}
The step size $s$ is chosen such that the acceptance rate is roughly 
$50 \%$. 
In the case of the local overrelaxation update, the new value of the
field is given by
\begin{equation}
\phi_x' =  2 \frac{\phi_x \cdot \Phi_x}{(\Phi_x)^2} \Phi_x -\phi_x
\end{equation}
where $\Phi_x = \sum_{y.nn.x} \phi_y$ is the sum over all fields that
live on sites $y$ that are nearest neighbors of $x$.  In the case of
the local updates we run through the lattice in typewriter
fashion. Going through the lattice once is called one sweep.
We use the following cycle of updates: One Metropolis sweep, one
overrelaxation sweep, $L/2$ single cluster updates, two overrelaxation
sweeps and finally $L/2$ single cluster updates. In this cycle, we
compute the observables after $L/2$ single cluster updates, i.e.
twice.

The average size of a cluster is proportional to the magnetic
susceptibility that grows like $L^{2-\eta}$.  Therefore, with our
choice of $L/2$ single cluster updates per cycle, the fraction of
sites that is updated by the cluster algorithm in one cycle of the
algorithm stays roughly constant.  We also note that the
overrelaxation update takes very little CPU time compared with the
Metropolis update. For $L=32$ and 
$N=3$ the CPU time
needed for one overrelaxation sweep, one Metropolis sweep, and $L/2$
single cluster updates roughly behave as $1:4:3$.

In all our simulations we have used the SIMD-oriented Fast Mersenne Twister
algorithm \cite{twister} as pseudo-random number generator.

\subsection{Statistics of the simulation}
\label{statsMC}

In the case of the XY universality class, we performed most of our
simulations at $\lambda= 2.1$ and $\beta=0.5091503$. We simulated the
lattice sizes $L=6,7,8,...,18$ and $20,22,24,26,28$. Throughout we
performed $10^9$ measurements.  In total these simulations took about
7 month of CPU time on a single core of a Quad-Core AMD Opteron(tm)
Processor 2378 running at 2.4 GHz.  In addition we performed
simulations at $\lambda=2.2$ and $\beta=0.508336$ where we simulated
the lattice sizes $L=6,7,8,...,12$. The results for $\lambda=2.2$
are used to estimate the effect of the uncertainty of $\lambda^*$.
Note that $\lambda^*=2.15(5)$ \cite{CHPV-06}. The values of $\beta$
chosen for the simulations at $\lambda=2.1$ and $2.2$ are 
the estimates of $\beta_c$ given in Table II of Ref.~\cite{CHPV-06}.

In the O(3) case we performed most simulations for $\lambda=4.5$ which
is close to our old estimate $\lambda^*=4.6(4)$ \cite{CHPRV-02}.  We
simulated at $\beta=0.686238$ which is close to the estimate
$\beta_c=0.6862385(20)$ \cite{CHPRV-02}. For the lattice sizes
$L=6,7,8,9,...,16$ we performed $10^9$ measurements, for
$L=17,18,...,32$ between $1.1 \times 10^9$ and $1.2 \times 10^9$
measurements and $5 \times 10^8$, $2.5 \times 10^8$, and $10^6$
measurements for $L=48$, $64$ and $256$, respectively.  In total these
simulations took about 4 years of CPU time on a single core of a
Quad-Core AMD Opteron(tm) Processor 2378 running at 2.4 GHz.  In
addition, we performed MC simulations at $\lambda=4.0$,
$\beta=0.68439$ and $\lambda=5.0$, $\beta=0.687564$ on lattices of the
size $L=6,7,8,...,16$.  Throughout we performed $10^9$
measurements. These results are used to determine our new estimate of
$\lambda^*$ and the effect of the uncertainty of $\lambda^*$ on our
estimates of the RG exponents.

In the O(4) case most of our simulations
were done for $\lambda=12.5$ and $\beta=0.9095167$.  For
$L=6,7,8,...,18$ and $20,22,24,26,28$ we performed $10^9$ measurements
and for $L=40$ we performed $6.5 \times 10^8$ measurements. 
For $L=256$ we performed $10^6$ measurements and simulated at
$\beta=0.909513$, which was our preliminary value of $\beta_c$. This
simulation was done to get a better estimate of $\beta_c$. In this
simulation we did not measure the quantities $C_l$ and $D_l$. 
The estimate $\beta=0.9095167$ used above was obtained by
requiring that $\xi/L=0.547$ which is the result for the large volume limit
$(\xi/L)^*$ of \cite{H-01}.
In addition, in order to determine $\lambda^*$ and the effect of the
uncertainty of $\lambda^*$ on the accuracy of our estimates of the
RG-exponents, we have simulated at $\lambda=14$ the lattice sizes
$L=6,7,8,...12$; $\lambda=18$ the lattice sizes $L=6,7,8,...12$;
$\lambda=22$ the lattice sizes $L=6,7,8,...16,18,20$; $\lambda=30$ and
$32$ the lattice size $L=6$; and for $\lambda=\infty$ the lattice sizes
$L=6,7,8,...12,16,24,32$. Throughout the statistics is $10^9$
measurements.

The  CPU time used for the whole study amounts to
roughly 7 years on a single core of a Quad-Core AMD Opteron(tm)
Processor 2378 running at 2.4 GHz.

\subsection{Variance of the observables}
\label{varobs}
The behavior of the variance of the quantities considered in our MC
simulations strongly affects the design of our study. The main
problem, as already observed in ref. \cite{CH-98} is that the relative
statistical error, at a fixed number of updates, of $C_4$ and $D_4$ rapidly
increases with the lattice size.  Therefore we have to focus on smaller
lattice sizes than one would do in a study mainly aiming at the
exponents $\nu$ and $\eta$.

Let us discuss this problem in a bit more detail at the example of the
simulations for $N=3$, $\lambda=4.5$ and the quantities $D_l$. 
Since we average over 10000 measurements at simulation time, we
can not disentangle integrated autocorrelation time and variance of
the quantities. Therefore in the following we discuss the relative
statistical error, normalized to $10^9$ measurements. In the case of
$D_4$ this relative statistical error is increasing from $0.000175$
for $L=6$ up to $0.051$ for $L=256$. This increase is well described
by a power law $e \propto L^x$, with $x \approx 1.45$.  Also in the
case of $D_3$ the relative error is increasing; $0.000064$ for $L=6$
up to $0.00022$ for $L=256$. However here the increase is smaller; it
is characterized by the exponent $x\approx 0.3$.  Interestingly, for
$D_2$ we find that the relative statistical error is even decreasing a
bit; $0.000037$ for $L=6$ down to $0.00003$ for $L=256$. The
corresponding exponent is $x\approx -0.05$.  This behavior can be
compared with that of the relative error of the slope of the Binder
cumulant or the second moment correlation length. These quantities are
used to determine the critical exponent $\nu$.  In both cases we find
a mild increase of the relative error, which is characterized by the
exponents $x\approx 0.06$ and $x\approx 0.14$, respectively.

As shown in Ref.~\cite{CH-98}, the problem of the large variance of
$C_4$ can be reduced by performing a larger number of
overrelaxation updates which are relatively cheap in terms of CPU
time and measure $C_4$ after each such update. This way one could
improve the efficiency in terms of 1/[(CPU-time)$\times$error$^2$]
of $C_4$ or $D_4$ by about a factor of 2
compared with the update cycle used in our simulations. However, since
this would have an 
adverse effect with respect to all other quantities
that we have measured we abstained from this.

For several observables, such as the susceptibility and the quartic
Binder cumulant, the statistical errors at fixed $\xi/L$ are smaller
than those at fixed $\beta$ close to $\beta_c$. Some comparisons are
reported in Refs.~\cite{CHPV-06,Parisen-11}. This is due to cross
correlations and to a reduction of the effective autocorrelation
times.  Taking $C_l$ or $D_l$ at $\xi/L$ fixed reduces the variance in
a $l$-dependent way.
For the $C_4$ and $D_4$ cases there is virtually no reduction of the
error. For $L=6$ there is still an improvement by a few percent,
however with increasing $L$, the ratio of errors goes rapidly to
$1$. In the $l=3$ case we observe a mild improvement by fixing
$\xi/L$.  For $C_3$ the ratio of statistical errors is $1.9$ for
$L=6$, $1.10$ for $L=64$ and $1.017$ for $L=256$.  In the case of
$D_3$, the ratio of statistical errors is $1.33$ for $L=6$, $1.06$
for $L=64$ and $1.014$ for $L=256$.  The reduction of the statistical
error is most significant in the $l=2$ case. For $C_2$ the ratio of
the statistical errors is $3.49$ for $L=6$, it decreases to $2.65$ at
$L=27$ and then increases again; $2.69$ at $L=64$ and $2.88$ for
$L=256$. For $D_2$ the ratio of the statistical errors is $2.21$ for
$L=6$, has its minimum $1.91$ at $L=23$, takes $2.02$ for $L=64$ and
$2.20$ for $L=256$.

\section{Some further results for the O($N$) vector models, $N=3$ and $4$}
\label{fures}

\subsection{New estimate for $\beta_c$}

In order to determine $\beta_c$, we fit the data for
$\xi/L$ and $U_4$ at $\lambda=4.5$ to the ansaetze
\begin{equation}
\label{R0}
 R(L,\beta_c) = R^* \;\;
\end{equation}
\begin{equation}
\label{R1}
 R(L,\beta_c) = R^*+ a L^{-0.79} 
\end{equation}
and 
\begin{equation}
\label{R2}
 R(L,\beta_c) = R^*+ a L^{-0.79} + b L^{-\epsilon}
\end{equation}
where either $\epsilon=1.6$ or $\epsilon=2$. Here we
take $0.79$ as value of the correction exponent $\omega$. 
By replacing it with $0.77$ say, 
our results for $\beta_c$ and $R^*$ change only very little.
In this study, 
we only calculate first derivatives of the quantities;
therefore in the fits we use the approximation 
\begin{equation}
\label{Taylor1}
 R(L,\beta) \approx R(L,\beta_s) + a (\beta-\beta_s)
\end{equation}
where $\beta_s$ is the value of the inverse temperature used
for the simulation. Since $\beta_s$ is very close to our final
result for $\beta_c$, the error due to the 
truncation of the Taylor-series can be ignored.

Let us first discuss the analysis of $\xi/L$. 
Taking no corrections into account, i.e. fitting with the
ansatz~(\ref{R0}), $\chi^2/$DOF remains unacceptably large
until most of our lattice sizes are discarded.
Including $L=48,64$ and $256$, we obtain
$(\xi/L)^*=0.56421(5)$, $\beta_c-\beta_s=-0.0000006(5)$
and $\chi^2/$DOF$=1.72/1$.  Using the ansatz~(\ref{R1}), i.e.
adding a correction term $a L^{-0.79}$ we get a $\chi^2/$DOF
smaller than 1 starting from $L_{min}= 12$, where all lattice
sizes $L \ge L_{min}$ are taken into account. Discarding 
further data points $\chi^2/$DOF is further decreasing and 
$(\xi/L)^*$ and $\beta_c-\beta_s$ move monotonically.
For $L_{min} = 18$ we find $(\xi/L)^*=0.56405(5)$ and 
$\beta_c-\beta_s=-0.00000067(38)$.  Adding a further correction, 
we get acceptable values of $\chi^2/$DOF  already for 
$L_{min}=7$. But also here  $\chi^2/$DOF still further decreases
and $(\xi_{2nd}/L)^*$ and $\beta_c-\beta_s$ move monotonically
with increasing $L_{min}$.
For $\epsilon=1.6$, we obtain  the results
$(\xi_{2nd}/L)^*=0.56386(10)$ and $\beta_c-\beta_s=-0.00000119(48)$
for $L_{min}=12$.
For $\epsilon=2$ and $L_{min}=12$, we get the results
$(\xi_{2nd}/L)^*=0.56391(8)$ and $\beta_c-\beta_s=-0.0000011(46)$.
For the Binder cumulant similar results can be found. We arrive
at the final results $\beta_c(\lambda=4.5) = 0.6862368(10)$ and
\begin{equation}
 (\xi/L)^* = 0.5639(2),\qquad U_4^* = 1.1394(3).
\end{equation}
The error-bars are chosen such that the results of the different fits
are covered.

A similar analysis for the O(4) symmetric $\phi^4$ model at
$\lambda=12.5$ leads to estimates $U_4^* = 1.0942(3)$, $\xi/L=
0.5471(3)$, and $\beta_c = 0.909517(2)$.

\subsection{Determination of $\lambda^*$}
\label{lambda*}

Next we determine 
the value of $\lambda^*$  where leading corrections to scaling vanish.
To this end we study 
\begin{equation}
 \bar U_4(L) = U_4(L,\beta_f)
\end{equation}
where $\beta_f$ is determined by the equation
\begin{equation}
\frac{\xi(L,\beta_f)}{L} = 0.5644
\end{equation}
where $0.5644$ is the result for $(\xi/L)^*$
of Ref. \cite{CHPRV-02}. In order to compute $\bar U_4$ we use 
the first order Taylor expansion~(\ref{Taylor1}) of $\xi/L$ and $U_4$
around the simulation point $\beta_s$. 
For $L=12$ , $\lambda=4.5$ we simulate at a number 
of different $\beta_s$, to check whether this approximation 
is sufficient for our purpose. In particular we find 
that for $\lambda=4.5$ the difference between $\beta_s=0.686238$
and $\beta_f$ is sufficiently small that contributions
$\propto (\beta - \beta_s)^2$ can be ignored. Due to scaling,
we expect that this also holds for all of the lattice 
sizes that we have simulated.

First we fit our data obtained at $\lambda=4.5$ with a number of
different ansaetze 
\begin{equation}
\label{fitU1}
\bar{U}_4 = \bar{U}_4^* + a L^{-0.79} \;\; ,
\end{equation}
\begin{equation}
\label{fitU2}
\bar{U}_4 = \bar{U}_4^* + a L^{-0.79}  + b L^{-\epsilon_1} \;\; ,
\end{equation}
and
\begin{equation}
\label{fitU3}
\bar{U}_4 = \bar{U}_4^* + a L^{-0.79}  + b L^{-\epsilon_1} + 
c L^{-\epsilon_2} \;\; .
\end{equation}
Also here we fix $\omega=0.79$; the final results change only little when we
replace it with $\omega=0.77$. In the case of the ansatz~(\ref{fitU2}) we set
$\epsilon_1=1.6$ or $2$. Finally in ansatz~(\ref{fitU3}) we add two terms with
subleading corrections.  We have fitted using various choices for $\epsilon_1$
and $\epsilon_2$.

In our fits we take into account all lattices sizes 
$L\ge L_{min}$.  In the case of the ansatz~(\ref{fitU1}) we get
an acceptable $\chi^2/$DOF starting from $L_{min}=22$. From this 
fit we get $a=0.00254(31)$. Further increasing $L_{min}$, 
$a$ is monotonically increasing; for $L_{min}=30$ we obtain
$a=0.0037(6)$.

Fitting with the ansatz~(\ref{fitU2}) and $\epsilon_1=1.6$ 
we obtain an acceptable 
$\chi^2/$DOF already starting from $L_{min}=6$. We get  
$a=0.01038(27)$ for the correction amplitude. Increasing 
$L_{min}$ the correction amplitude remains stable.
Using instead $\epsilon_1=2$  we get an
acceptable $\chi^2/$DOF starting from $L_{min}=7$.  The corresponding 
result for the correction amplitude is $a=0.00586(20)$. Increasing $L_{min}$,
the value of $a$ increases up to $a=0.00676(33)$ for $L_{min}=10$.
For $L_{min}=11$ and $12$ we get a very similar result. 
For $L_{min}=12$,  $\chi^2/$DOF $=14.50/21$ and $15.78/21$ for 
$\epsilon=2$ and $1.6$, respectively.

Finally we fit with the ansatz~(\ref{fitU2}) using
$(\epsilon_1,\epsilon_2)=(1.6,2)$, $(1.6,1.96)$ or $(1.8,2)$.  The results of
such fits are all in the interval $0.005 < a < 0.011$.  We conclude $a =
0.007(4)$, where the central value and the error-bar are chosen such that the
results of the different fits are covered. Next we convert this estimate of
the correction amplitude at $\lambda=4.5$ into a new estimate of $\lambda^*$.
In order to compute the derivative of $a$ with respect to $\lambda$, we study
the differences
\begin{equation}
 \Delta \bar{U}_4(L) = \bar{U}_4(L,\lambda=5) - \bar{U}_4(L,\lambda=4) \;\;.
\end{equation}
In this difference $\bar U_4^*$ exactly cancels. Furthermore subleading 
corrections should cancel to a large extend. Therefore we fit
our data with the ansatz
\begin{equation}
\label{deltafit}
 \Delta \bar{U}_4(L) = c L^{-\omega} \;\;.
\end{equation}
Results of such fits with  $c$ and $\omega$ as free parameters 
are given in table~\ref{tabomega}. 
\begin{table}[tbp]
\caption{Fits with the ansatz~(\ref{deltafit}), O(3) universality class
}
\label{tabomega}
\begin{ruledtabular}
\begin{tabular}{cccc}
$L_{min}$ &  $c$   &    $\omega$  &    $\chi^2/$DOF \\
\hline
 6 &  -0.0109(2) &  0.775(9)\phantom{0}  &  6.44/9  \\
 7 &  -0.0109(3) &  0.777(12) &  6.38/8  \\
 8 &  -0.0111(4) &  0.784(16) &  5.86/7  \\
\end{tabular}
\end{ruledtabular}
\end{table}
Already starting  from $L_{min}=6$ we get an acceptable 
$\chi^2/$DOF. Furthermore, the value obtained 
for $\omega$ is fully consistent with the field 
theoretic estimates $\omega=0.782(13)$ and 
$\omega=0.794(18)$ obtained by the perturbative 
expansion in three dimensions fixed and the $\epsilon$-
expansion, respectively \cite{GZ-98}.  
The facts that $\chi^2/$DOF
is small and the result for $\omega$
is consistent with the field-theoretical ones
confirm our assumption that already
for the lattice sizes that we consider, $\Delta \bar{U}_4(L)$
is dominated by the leading correction.

Fitting with $\omega=0.79$ fixed, to be consistent with 
the analysis of $\bar U_4$ at $\lambda=4.5$ above, we find 
$c=-0.01126(4)$ and $\chi^2/$DOF$=6.0/8$ for $L_{min}=8$.
The result for $c$ changes little, when $L_{min}$ is
varied. In order to check how well the derivative of 
$a$ with respect to $\lambda$ is approximated by the 
finite difference, we also have fitted  
$\bar{U}_4(L,\lambda=5) - \bar{U}_4(L,\lambda=4.5)$. 
Here we find $c=-0.00506(4)$ and $\chi^2/$DOF$=5.4/8$
for $L_{min}=8$. Also here, the result for $c$ changes little, 
when $L_{min}$ is varied. 

Using these results we arrive at 
\begin{equation}
 \lambda^* \approx 4.5 - a(\lambda=4.5) 
\left( \frac{\partial a}{\partial \lambda} \right)^{-1} 
   = 4.5 - 0.007(4)/(-2 \times 0.00506(4) ) \approx 5.2(4) \;\;.
\end{equation}

We perform a similar analysis in the case of the O(4) 
universality class. Here $\beta_f$ is given by
\begin{equation}
\frac{\xi(L,\beta_f)}{L} = 0.547
\end{equation}
where $0.547$ is the result for $(\xi/L)^*$
of ref. \cite{H-01}.  First we have analyzed the data for $\bar U_4$
at $\lambda=12.5$. The analysis is done in much the same way as
discussed above in detail for the O(3) universality class.
Fixing $\omega=0.79$ we find $a = 0.007(5)$ as amplitude of the
leading correction. 

Next we study  the difference 
\begin{equation}
 \Delta \bar{U}_4 (L,\lambda_1,\lambda_2) = 
\bar{U}_4 (L,\lambda_1)-\bar{U}_4 (L,\lambda_2) \;\;.
\end{equation}
We perform fits for $\lambda_1 = 22$, $\lambda_2=12.5$ and
$\lambda_1 = \infty$, $\lambda_2=12.5$ using the ansatz~(\ref{deltafit}) 
with $c$ and $\omega$ as free parameters. The results for 
$\lambda_1 = 22$  and $\lambda_1 = \infty$ are given
in tables \ref{tabomega2} and \ref{tabomega4}, respectively.

\begin{table}[tbp]
\caption{Fits of $\Delta \bar{U}_4 (L,22,12.5)$
with the ansatz~(\ref{deltafit}), O(4) universality class.
}
\label{tabomega2}
\begin{ruledtabular}
\begin{tabular}{cccc}
$L_{min}$ &  $c$   &    $\omega$  &    $\chi^2/$DOF \\
\hline
 6 & -0.00776(14) & 0.777(8)\phantom{0} & 6.77/11   \\
 7 & -0.00764(18) & 0.771(10) & 5.81/10 \\
 8 & -0.00753(22) & 0.765(11) & 5.15/9 \\
 9 & -0.00741(28) & 0.759(15) & 4.64/8 \\
\end{tabular}
\end{ruledtabular}
\end{table}

\begin{table}[tbp]
\caption{Fits of $\Delta \bar{U}_4 (L,\infty,12.5)$
with the ansatz~(\ref{deltafit}), O(4) universality class.
}
\label{tabomega4}
\begin{ruledtabular}
\begin{tabular}{cccc}
$L_{min}$ &  $c$   &    $\omega$  &    $\chi^2/$DOF \\
\hline
 6 & -0.01870(16) & 0.787(4) & 11.64/7  \\
 7 & -0.01849(21) & 0.783(5) & 9.39/6  \\
 8 & -0.01841(26) & 0.781(6) & 9.05/5  \\
 9 & -0.01844(32) & 0.782(7) & 9.02/4  \\
10 & -0.01846(38) & 0.782(8) & 9.00/3  \\
11 & -0.01777(45) & 0.769(10)& 2.03/2  \\
\end{tabular}
\end{ruledtabular}
\end{table}
These results can be compared with $\omega=0.774(20)$ and 
 $\omega=0.795(30)$ from the perturbative expansion at
three dimensions fixed and the $\epsilon$-expansion, respectively 
\cite{GZ-98}. 

Fixing $\omega=0.79$ we obtain $c=-0.00800(2)$ (with $\chi^2/$DOF=9.77/12) as
amplitude for the differences $\lambda_1 = 22$ and $\lambda_2=12.5$ with
$L_{min}=6$ Taking data only for $L=6$ we get $c(\lambda_1=14,12.5)
=-0.00193(5)$ $c(\lambda_1=20,12.5) =-0.00696(5)$, $c(\lambda_1=30,12.5)
=-0.01084(5)$ $c(\lambda_1=32,12.5) =-0.01132(5)$. It is quite clear from
these numbers that a linearization of the correction amplitude as a function
of $\lambda$ is not sufficient to compute the estimate of $\lambda^*$. For the
same reason, we give an asymmetric estimate of the error:
\begin{equation}
\lambda^* = 20_{-6}^{+15}
\end{equation}
This value is larger than $\lambda^* =12.5(4.0)$ that we quote in
ref. \cite{H-01}. However we are quite confident that indeed a 
$\lambda^*$ exists for the O(4) 
case.  Note that in the limit $N \rightarrow \infty$ for the simple
cubic lattice and the given  lattice action, no $\lambda^*$ exists 
and that leading corrections are minimal 
in the limit $\lambda \rightarrow \infty$ \cite{ZJ-book}. 

\subsection{The magnetic susceptibility and the exponent $\eta$}
In order to obtain the critical exponent $\eta$, we analyze the 
behavior of 
\begin{equation}
\bar \chi = \chi(\beta_f)
\end{equation}
where, in the O(3) case  $\beta_f$ 
is defined by $\xi(\beta_f)/L=0.5644$. In the first step
of the analysis we eliminate leading corrections to scaling.
To this end we analyze the ratios
\begin{equation}
\frac{\bar{\chi}(\lambda=5)}{\bar{\chi}(\lambda=4)} 
  = a (1+ c L^{-0.79}) \;\;.
\end{equation}
We obtain a good fit starting from $L_{min}=11$. For $L_{min}=11$ we
obtain $a=0.99172(8)$, $c=-0.0046(6)$ and $\chi^2/$DOF$=3.11/4$.
Therefore in order to eliminate corrections at $\lambda=4.5$ we follow
the strategy discussed in section \ref{rescorr}.  Using
$\bar{U}_4=U_4^* + 0.007(4) L^{-0.79} + ...$ and $\bar{U}_4(\lambda=5)
- \bar{U}_4(\lambda=4) = -0.01126(4) L^{-0.79} ...$)
Eq.~(\ref{tildecorrect}) reads
\begin{equation}
{\widetilde \chi} \equiv \bar{\chi}(\lambda=4.5) 
\left (1- \frac{-0.0046(6)}{-0.01126(4)} 0.007(4) L^{-0.79} \right)
\end{equation}
We fit $\widetilde{\chi}$ with the ansaetze 
\begin{equation}
\label{chi0}
 \widetilde{\chi} = a L^{2-\eta}  \;\;,
\end{equation}
\begin{equation}
\label{chi1}
 \widetilde{\chi} = a L^{2-\eta}  + c \;\;,
\end{equation}
\begin{equation}
\label{chi2}
 \widetilde{\chi} = a L^{2-\eta} (1 +b L^{-\epsilon})  + c
\end{equation}
with $\epsilon=1.6$ or $\epsilon=1.8$.  In the case of the
ansatz~(\ref{chi0}) we obtain very large $\chi^2/$DOF up to
$L_{min}=32$. For $L_{min}=48$ we get $\eta=0.0375(1)$ and
$\chi^2/$DOF$=0.46/1$.  Using the ansatz ~(\ref{chi1}) we get
$\chi^2/$DOF$\approx 1$ already for $L_{min}=16$; for example, for
$L_{min}=18$ we obtain $\eta=0.03767(4)$ and $\chi^2/$DOF$=10.11/15$.
Using the ansatz~(\ref{chi2}) with $\epsilon=1.6$ we get for
$L_{min}=10$ the results $\eta=0.03791(7)$ and
$\chi^2/$DOF$=14.55/22$.  and for $\epsilon=1.8$ and $L_{min}=8$ we
get $\eta=0.03780(3)$ and $\chi^2/$DOF$=18.74/24$.  We redo
these fits for $\bar{\chi}$ without correction to check the effect of
the uncertainty of $\lambda^*$.  We find that the estimates of $\eta$
change by about $0.0001$.  Taking into account only fits with ansaetze
that include the analytic background, we arrive at
\begin{equation}
\eta = 0.0378(3) \;\;.
\end{equation}

In the case of the O(4) universality class, performing a similar analysis 
we obtain
\begin{equation}
\eta = 0.0360(3) \;\;.
\end{equation}

\subsection{The exponent $\nu$}
We estimate the exponent $\nu$ from the behavior of the 
slope of $U_4$ and $\xi/L$ at $\beta_c$:
\begin{equation}
\label{Sbehavior}
 S_R = \left . \frac{\partial R}{\partial \beta} 
\right |_{\beta=\beta_c} = a L^{1/\nu}  
(1 + c L^{-\omega} +...).
\end{equation}
Since we did not plan to compute the exponent $\nu$ from the
beginning, we did not compute the second derivatives of $U_4$ and
$\xi/L$ with respect to $\beta$.  Hence we can not compute the
slope at fixed values of $U_4$ or $\xi/L$.  At $\lambda=4.5$ we
performed MC simulation very close to our final value of
$\beta_c$. Therefore it is sufficient to have a rather rough estimate
of the second derivatives of $U_4$ and $\xi/L$ in order to
compute the first derivatives of $U_4$ and $\xi/L$ at $\beta_c$
starting from the first derivatives of $U_4$ and $\xi/L$ at
$\beta_s$ that we have computed in our simulations.  To this end, we
simulated for $L=12$ at a number of different $\beta$
values. Using these data we compute the second derivatives of
$U_4$ and $\xi/L$ with respect to $\beta$ by finite
differences. The second derivatives are then estimated by $R'' (L) =
R''(12) (L/12)^{2/\nu}$. Notice that our estimate of
$\beta_c=0.6862368(10)$ is very close to the simulation point
$\beta_s=0.686238$.  We analyze the resulting data by fitting
with various ansaetze that are derived from Eq.~(\ref{Sbehavior}).  We
arrive at $\nu=0.7118(7)$ from the analysis of the slope of of $\xi/L$
and $\nu=0.7114(11)$ from that of $U_4$. The error bars take also into
account the uncertainty of $\lambda^*$.  As our final estimate we
quote
\begin{equation}
 \nu=0.7116(10)  \;\;.
\end{equation}

By a similar analysis for the O(4) universality  class, we obtain 
\begin{equation}
 \nu=0.750(2) \;\;.
\end{equation}


\begin{thebibliography}{99}

\bibitem{ZJ-book}
J.~Zinn-Justin, {\em Quantum Field Theory and Critical Phenomena},
third edition (Clarendon Press, Oxford, 1996).

\bibitem{PV-r}
A. Pelissetto and E. Vicari, Phys. Rep. 368, 549 (2002).

\bibitem{CHPV-06} M. Campostrini, M. Hasenbusch, A. Pelissetto, and
E. Vicari, Phys. Rev. B 74, 144506 (2006).

\bibitem{GZ-98}
R.~Guida and J.~Zinn-Justin, J. Phys. A 31, 8103 (1998).

\bibitem{NR-84}
K.E. Newman and E.K. Riedel, Phys. Rev. B 30, 6615 (1984).

\bibitem{CHPRV-02}
M. Campostrini, M. Hasenbusch, A. Pelissetto, P. Rossi, and E. Vicari,
Phys. Rev. B 65, 144520 (2002).

\bibitem{this} We use the new estimate $\lambda^*=5.2(4)$ to update
the high-temperature results of Ref.~\cite{CHPRV-02}, see its
Eqs.~(14) and (19).


\bibitem{H-01}
M.~Hasenbusch,  J. Phys. A  34, 8221 (2001).

\bibitem{Deng-06}
Y. Deng, Phys. Rev. E 73, 056116 (2006).

\bibitem{HPV-05}
M. Hasenbusch, A. Pelissetto, and E. Vicari,
Phys. Rev. B 72, 014532 (2005).

\bibitem{CPV-03}
P. Calabrese, A. Pelissetto, and E. Vicari, Phys. Rev. B 67, 054505 (2003).

\bibitem{Aharony-76} 
A.~Aharony, in {\em Phase Transitions and Critical Phenomena},
edited by C.~Domb and J.~Lebowitz 
(Academic Press, New York, 1976),
Vol.\ 6, p. 357.

\bibitem{FN-74} M.E. Fisher and D.R. Nelson, Phys. Rev. Lett. 32, 1350
(1974).

\bibitem{NKF-74} 
D.R. Nelson, J.M. Kosterlitz, and M.E. Fisher, 
Phys. Rev. Lett. 33, 13 (1974);
J.M. Kosterlitz, D.R. Nelson, and M.E. Fisher, Phys. Rev. B
13, 412 (1976).


\bibitem{Brock-etal-86}
J. D. Brock, A. Aharony, R. J. Birgeneau, 
K. W. Evans-Lutterodt, J. D. Litster,
P. M. Horn, G. B. Stephenson, and A. R. Tajbakhsh,
Phys. Rev. Lett. 57, 98 (1985).

\bibitem{ABBL-86}
A. Aharony, R. J. Birgeneau, J. D. Brock, and J. D. Litster,
Phys. Rev. Lett. 57, 1012 (1986).

\bibitem{Aharony-etal-95}
A. Aharony, R. J. Birgeneau, C. W. Garland, Y.-J. Kim,
V. V. Lebedev, R. R. Netz, and M. J. Young,
Phys. Rev. Lett. 74, 5064 (1995). 

\bibitem{AM-83}
S. R. Andrews and H. Mashiyama,
J. Phys. C 16, 4985 (1983).

\bibitem{HHTG-95}
G. Helgesen, J. P. Hill, T. R. Thurston, and D. Gibbs,
Phys. Rev. B 52, 9446 (1995).

\bibitem{ZMHCLGGS-96}
M. P. Zinkin, D. F. McMorrow, J. P. Hill, R. A. Cowley,
J.-G. Lussier, A. Gibaud, G. Gr\"ubel, and C. Sutter,
Phys. Rev. B 54, 3115 (1996).

\bibitem{Bak-80}
P. Bak, Phys. Rev. Lett. 44, 889 (1980).

\bibitem{Wegner-76} 
F.~J.~Wegner, in
{\em Phase Transitions and Critical Phenomena},
edited by C.~Domb and M.~S.~Green 
(Academic Press, New York, 1976), Vol.\ 6.

\bibitem{Nicoll-81}
J.F. Nicoll, Phys. Rev. A 24, 2203 (1981).

\bibitem{CH-98}
M. Caselle and M. Hasenbusch, J. Phys. A 31, 4603 (1998).

\bibitem{DPV-03}
M. De Prato, A. Pelissetto, and E. Vicari, Phys. Rev. B 68, 092403 (2003).

\bibitem{CPV-00}
J.M. Carmona, A. Pelissetto, and E. Vicari, Phys. Rev. B 61, 15136 (2000).

\bibitem{PJF-74}
P. Pfeuty, D. Jasnow, and M.E. Fisher, Phys. Rev. B 10, 2088 (1974).

\bibitem{ChFiNi}
J.H. Chen, M. E. Fisher, and B. G. Nickel, Phys.\ Rev.\ Lett.\ 48,
630 (1982);
M. E. Fisher and J. H. Chen, J. Physique (Paris) 46, 1645 (1985).

\bibitem{CHPRV-01}
M.~Campostrini, M.~Hasenbusch, A.~Pelissetto, P.~Rossi,
and E.~Vicari, Phys.\ Rev.\ B 63,  214503 (2001).

\bibitem{HT-99}
M.~Hasenbusch and T.~T\"or\"ok,  J.~Phys. A 32, 6361 (1999)

\bibitem{CPRV-99}
M.~Campostrini, A.~Pelissetto, P.~Rossi,
and E.~Vicari, Phys.\ Rev.\ E 60,  3526 (1999).

\bibitem{HPPV-07}
M. Hasenbusch, F. Parisen Toldin, A. Pelissetto, and E. Vicari, 
J. Stat. Mech.: Theory Exp. (2007) P02016.

\bibitem{CPV-02}
P. Calabrese, A. Pelissetto, and E. Vicari, 
Phys. Rev. B 65, 046115 (2002).

\bibitem{Brock-et-al-89}
J.D. Brock, D.Y. Noh, B.R. McClain, J.D. Lister,
R.J. Birgeneau, A. Aharony, P.M. Horn, J.C. Liang,
Z. Phys. B 74, 197 (1989).

\bibitem{Chikazumi-book}
S. Chikazumi, {\em Physics of Ferromagnetism}
(Clarendon, Oxford, 1997) Chapt. 12.

\bibitem{PS-00}
D.V. Pakhnin and A.I. Sokolov,
Phys. Rev. B  61, 15130   (2000).

\bibitem{FHY-00}
R. Folk, Yu. Holovatch, and T. Yavors'kii,
Phys. Rev. B  62, 12195 (2000);
(E) B  63, 189901 (2001).

\bibitem{KS-95}
H. Kleinert and V. Schulte-Frohlinde,
Phys. Lett. B 342, 284 (1995).

\bibitem{RG-77}
H. Rohrer and Ch. Gerber, Phys. Rev. Lett. 38, 909 (1977).

\bibitem{KR-79}
A.R. King and H. Rohrer, Phys. Rev. B 19, 5864 (1979).

\bibitem{OPSB-78}
N.F. Oliveira Jr., A. Paduan Filho, S.R. Salinas,
and C.C. Becerra, Phys. Rev. B 18, 6165 (1978). 

\bibitem{FHM-08}
R. Folk, Yu. Holovatch, and G.Moser,
Phys. Rev. E 78, 041124 (2008).

\bibitem{Selke-11}
W. Selke, Phys. Rev. E 83, 042102 (2011). 


\bibitem{Wolff}
U. Wolff, Phys. Rev. Lett. 62, 361 (1989).

\bibitem{twister}
M. Saito and M. Matsumoto,
``SIMD-oriented Fast Mersenne Twister:
a 128-bit Pseudorandom Number Generator'',
in
{\sl Monte Carlo and Quasi-Monte Carlo Methods 2006},
edited by A. Keller, S. Heinrich, H. Niederreiter, (Springer, 2008);
M. Saito, Masters thesis, Math. Dept., Graduate School of schience,
Hiroshima University, 2007.
The source code of the program is provided at
``http://www.math.sci.hiroshima-u.ac.jp/$\sim$m-mat/MT/SFMT/index.html''

\bibitem{Parisen-11}
F. Parisen Toldin, arXiv:1104.2500 

\end{thebibliography}
\end{document}